\catcode`\@=11

\newif\if@fewtab\@fewtabtrue

{\count255=\time\divide\count255 by 60
\xdef\hourmin{\number\count255}
\multiply\count255 by-60\advance\count255 by\time
\xdef\hourmin{\hourmin:\ifnum\count255<10 0\fi\the\count255}}
\def\ps@draft{\let\@mkboth\@gobbletwo
    \def\@oddhead{}
    \def\@oddfoot
       {\hbox to 7 cm{$\scriptstyle Draft\ version:\ \draftdate$
       \hfil}\hskip -7cm\hfil\rm\thepage \hfil}
    \def\@evenhead{}\let\@evenfoot\@oddfoot}

\def\ceqno{\global\@fewtabfalse
    \ifcase\@eqcnt \def\@tempa{& & &}\or \def\@tempa{& &}
      \or \def\@tempa{&}
      \or\def\@tempa{}\fi\@tempa
{\rm(\theequation)}}

\def\aeqno#1{\global\@fewtabfalse
    \ifcase\@eqcnt \def\@tempa{& & &}\or \def\@tempa{& &}
      \or \def\@tempa{&}
      \or\def\@tempa{}\fi\@tempa
{\rm(\theequation,#1)}}

\def\label#1{\ifnum\draftcontrol=1
 \global\def\draftnote{$\scriptstyle #1$}\fi
 \@bsphack\if@filesw {\let\thepage\relax
   \def\protect{\noexpand\noexpand\noexpand}%
\xdef\@gtempa{\write\@auxout{\string
      \newlabel{#1}{{\@currentlabel}{\thepage}}}}}\@gtempa
   \if@nobreak \ifvmode\nobreak\fi\fi\fi
  \@esphack}

\def\alabel#1#2{\label{#1}\global\@fewtabfalse
    \ifcase\@eqcnt \def\@tempa{& & &}\or \def\@tempa{& &}
      \or \def\@tempa{&}
      \or\def\@tempa{}\fi\@tempa
{\hbox to 3cm{\phantom{\rm(\theequation,#2)}
\draftnote \hfil}\hskip -3cm {\rm(\theequation,#2)}}}

\def\clabel#1{\label{#1}\global\@fewtabfalse
    \ifcase\@eqcnt \def\@tempa{& & &}\or \def\@tempa{& &}
      \or \def\@tempa{&}
      \or\def\@tempa{}\fi\@tempa
{\hbox to 3cm{\phantom{\rm(\theequation)}
\draftnote \hfil}\hskip -3cm{\rm(\theequation)}}}

\def\eqnarray{\def\draftnote{{}}\global\@fewtabtrue
\stepcounter{equation}\let\@currentlabel=\theequation
\global\@eqnswtrue
\global\@eqcnt\z@\tabskip\@centering\let\\=\@eqncr
$$\halign to \displaywidth\bgroup\@eqnsel\hskip\@centering\@eqcnt\z@
  $\displaystyle\tabskip\z@{##}$&\global\@eqcnt\@ne
  \hskip 1\arraycolsep \hfil${##}$\hfil
  &\global\@eqcnt\tw@ \hskip 1\arraycolsep
$\displaystyle\tabskip\z@{##}$
\hfil  \tabskip\@centering&\global\@eqcnt\thr@@\llap{##}\tabskip\z@
\cr}

\def\endeqnarray{\@@eqncr\egroup
      \global\advance\c@equation\m@ne$$\global\@ignoretrue}

\def\@eqnnum{\hbox to 3cm{\phantom{\rm(\theequation)} \draftnote
                         \hfil}\hskip -3cm {\rm(\theequation)}}

\def\@@eqncr{\let\@tempa\relax
    \ifcase\@eqcnt \def\@tempa{& & &}\or \def\@tempa{& &}
      \or \def\@tempa{&}
      \or\def\@tempa{}
\fi\@tempa
\if@eqnsw
\if@fewtab\@eqnnum\fi
\stepcounter{equation}\fi\global
\@eqnswtrue\global\@eqcnt\z@\global\@fewtabtrue\cr}
\def\erf#1{(\ref{#1})}

\def\draftcite#1{\ifnum\draftcontrol=1#1\else{}\fi}

\def\@lbibitem[#1]#2{\item{}\hskip -3cm \hbox to 2cm
{\hfil$\scriptstyle\draftcite{#2}$}\hskip
1cm[\@biblabel{#1}]\if@filesw
     {\def\protect##1{\string ##1\space}\immediate
      \write\@auxout{\string\bibcite{#2}{#1}}}\fi\ignorespaces}

\def\@bibitem#1{\item\hskip -3cm \hbox to 2cm
{\hfil $\scriptstyle\draftcite{#1}$}\hskip 1cm
\if@filesw \immediate\write\@auxout
       {\string\bibcite{#1}{\the\value{\@listctr}}}\fi\ignorespaces}

\font\tendl=msbm10  scaled \magstep1%double line
\font\sevendl=msbm7 scaled \magstep1
\font\fivedl=msbm5 scaled \magstep1
\font\tengl=eufm10  scaled \magstep1% gothic letters
\font\sevengl=eufm7 scaled \magstep1
\font\fivegl=eufm5 scaled \magstep1

\newfam\dlfam  % \dl is double line
\textfont\dlfam=\tendl \scriptfont\dlfam=\sevendl
\scriptscriptfont\dlfam=\fivedl
\newfam\glfam  % \gl is gothic letters
\textfont\glfam=\tengl \scriptfont\glfam=\sevengl
\scriptscriptfont\glfam=\fivegl

\def\draftdate{\number\month/\number\day/\number\year\ \ \ \hourmin }

\global\def\draftcontrol{0}
\catcode`\@=12

\documentclass[aps]{revtex4}

\usepackage{epsfig}
\usepackage[all]{xy}
\usepackage[centerlast]{subfigure}
\usepackage{bm}
\usepackage{amssymb}
\usepackage{amsmath}
\usepackage{amstext}
\renewcommand{\theequation}{\arabic{equation}}

\newcommand{\be}{\begin{eqnarray}}
\newcommand{\en}{\end{eqnarray}\vs 0.5 cm}

\newcommand{\no}{\noindent}
\newcommand{\vs}{\vskip}

\newcommand{\Ng}{{\bf g}}

\newcommand{\NR}{{{\bf R}}}%letra doble raya en modo matematico
\newcommand{\NT}{{{\bf T}}}%letra doble raya en modo matematico
\newcommand{\NZ}{{{\bf Z}}}%letra doble raya en modo matematico
\newcommand{\qq}{\begin{eqnarray}}

\newcommand{\ee}{{\rm e}}

\newcommand{\qqq}{\end{eqnarray}}

\newcommand{\tr}{\hbox{tr}}

\newcommand{\CG}{{\cal G}}
\newcommand{\CH}{{\cal H}}
\newcommand{\CI}{{\cal I}}

\newcommand{\CK}{{\cal K}}

\newcommand{\CM}{{\cal M}}

\newcommand{\CP}{{\cal P}}

\def\id{\mathrm{id}}
\newcommand{\alxydim}[2]{\begin{aligned}\xymatrix#1{#2}\end{aligned}}

\begin{document}

\title{{\Large\bf{Polyakov-Wiegmann Formula and Multiplicative Gerbes}}}

\author{Krzysztof Gaw\c{e}dzki}
\affiliation{Laboratoire de Physique, C.N.R.S., ENS-Lyon,\\
Universit\'e de Lyon, 46 All\'ee d'Italie, 69364 Lyon, France}

\author{Konrad Waldorf}
\affiliation{Department of Mathematics, University of California, Berkeley, 970 Evans Hall \#3840, Berkeley, CA 94720, USA}

\begin{abstract}
\no An unambiguous definition of Feynman amplitudes in  
the Wess-Zumino-Witten sigma model and the Chern-Simon gauge theory  
with a general Lie group 
is determined by a certain geometric 
structure on the group. For the WZW amplitudes, this is a (bundle) gerbe 
with connection of an appropriate curvature whereas for the CS amplitudes, 
the gerbe has to be additionally equipped with a multiplicative structure 
assuring its compatibility with the group multiplication. We show that 
for simple compact Lie groups
the obstruction to the existence of a multiplicative structure is provided 
by a 2-cocycle of phases that appears in   
the Polyakov-Wiegmann formula  relating the Wess-Zumino action 
functional of the product of group-valued fields to the sum of the 
individual contributions. 
These phases were computed long time ago for all compact simple Lie groups.
If they are trivial, then the multiplicative 
structure exists and is unique up to isomorphism.
\end{abstract}

\maketitle

\section{Introduction}
\label{sec:intro}

\no It has been known from the early works \cite{Witt,GepWitt,AbGep,FGK}
on the Wess-Zumino-Witten (WZW) two-dimensional sigma models that the 
consistency of such quantum field theories imposes restrictions on 
the possible values of the coupling constant $\,k\,$ called the level. 
\,In the more modern geometric language, the consistency requires 
the existence of a  gerbe with  connection over the target group $\,G$,
with the curvature of the gerbe equal to the closed 3-form
\qq
H_k\ =\ \frac{k}{24\pi^2}\,\tr\,(g^{-1}dg)^3
\label{meg}
\qqq
on $\,G\,$ \cite{CMM}\cite{top}\cite{G05}\cite{GR02}. 
Such a  gerbe $\,\CG_k\,$ exists if and only if the periods of 
$\,H_k\,$ are integers. \,For simple compact simply-connected groups,
this occurs when $\,k\in\NZ$, \,assuming a proper normalization of 
the bilinear $ad$-invariant form $\,\tr\,XY\,$ on the Lie algebra 
$\,\Ng\,$ of $\,G\,$ that appears on the right hand side of Eq.\,(\ref{meg}).
For non-simply connected groups, the integrality of the periods of 
$\,H_k\,$ may impose more constraints on the level $k$. \,For example, 
the consistency of the WZW model with the $\,SO(3)\,$ target requires 
even levels. In \cite{FGK}, such restrictions were analyzed for all 
simple compact groups. Similar results were obtained in \cite{SY,SY1,GRS,KS} 
via an algebraic approach that interpreted the corresponding WZW models as 
``simple current orbifolds''. 
\vskip 0.1cm

The gerbe $\,\CG_k\,$ over  $\,G\,$ determines in a canonical way 
the ``holonomy''
\qq
{\CH}_{\CG_k}(\varphi)\ \in\ U(1)
\nonumber
\qqq
defined for maps $\,\varphi\,$ from a closed oriented 
surface $\,\Sigma\,$ to $\,G\,$ \cite{top,CMM,GR02}. \,By definition, such maps
are the classical fields of the WZW model and the holonomy 
$\,{\CH}_{\CG_k}(\varphi)\,$ defines the contribution of the Wess-Zumino 
action to the Feynman amplitude of the field $\varphi$. 
\,The gerbe holonomy is invariant under the composition of fields with
orientation-preserving diffeomorphisms $\,D\,$ of $\,\Sigma$:
\qq
{\CH}_{\CG_k}(\varphi)\ = \ {\CH}_{\CG_k}(\varphi\circ D)\,.
\label{diffinv}
\qqq
The other important property of the holonomy relating it to the curvature 
form of the gerbe is the identity 
\qq
{\CH}_{\CG_k}(\varphi_1)\ 
=\ {\CH}_{\CG_k}(\varphi_0)\ \exp\Big[2\pi i\hspace{-0.3cm}
\int\limits_{[0,1]\times \Sigma}\hspace{-0.3cm}\phi^*H_k\Big]
\label{holcurv}
\qqq
holding for 1-parameter families (i.e. homotopies) of classical fields 
$\,\varphi_t=\phi(t,\,\cdot\,)\,$ with $\ \phi:[0,1]\times\Sigma
\rightarrow G$. 
\vskip 0.1cm

As noticed in \cite{WittCS}, the 2-dimensional WZW theory with 
simply-connected target groups is closely related to the 3-dimensional 
Chern-Simons (CS) gauge theory of the same level $\,k$. \,The existence 
of the CS theory with a non-simply connected gauge group imposes, however, 
stronger restrictions on the level \cite{ZOO,G90}. For example, the 
$\,SO(3)\,$ CS theory requires $\,k\,$ divisible by $\,4$. \,The topological 
origin of the difference between the two restrictions has been explained 
in \cite{DijkWitt}. \,In \cite{CJMSW}, the cohomological discussion of
\cite{DijkWitt} was lifted to the geometric level by showing that 
the CS theory with gauge group $\,G\,$ requires an additional structure 
on the gerbe $\,\CG_k\,$ 
turning it into a ``multiplicative gerbe''. 
The argument was completed in \cite{WMult} by including connections 
into the discussion of multiplicative structures.
\,It was shown there that a multiplicative gerbe $\,\CG_k\,$ with 
connection permits to define unambiguously Feynman amplitudes 
of the CS theory.
%KG: I restored what you didn't want. It seems to me that it elucidates 
%    the most important use of the multiplicative gerbes for the reader 
%    unfamiliar with [4] and [31] which otherwise may seem very mysterious. 
%    You may throw it out again if you really do not like it.
More exactly, for every gauge connection $\,A\,$ on a
        $\,G$-bundle over a manifold $\,M$, \,it determines canonically
        a 2-gerbe $\,\CK(A)\,$ over $\,M\,$ (a geometric structure of one
        degree higher) with curvature equal to the Pontryagin 4-form
        $\,\frac{k}{8\pi^2}\,\tr\,F(A)^2$. \,Given a map $\,\phi\,$ of
a        closed oriented 3-dimensional manifold into $\,M$, \,the CS Feynman
        amplitude of the gauge field $\,A\,$ is given as the holonomy of
        the 2-gerbe $\,\CK(A)\,$ along $\,\phi\,$ \cite{WMult}.
It was also shown in \cite{WMult} that the multiplicative gerbe 
$\,\CG_k\,$ 
determines canonically a central extension of the loop 
group $\,LG$. \,The latter provides the extended chiral algebra 
of the corresponding WZW theory whereas the WZW models corresponding
to gerbes $\,\CG_k\,$ without multiplicative structure 
possess less extended or unextended chiral algebras. 
%KG: I changed the wording:
\,E.g. the chiral
algebra of the $\,SO(3)\,$ WZW theory with $k$ divisible by 4 is the
provided by the central extension of $LSO(3)$ whereas for $\,k\,$ even 
but not divisible by $\,4\,$ it is given by the central extension 
of the loop group $\,LSU(2)\,$ \cite{KS}. 
\vskip 0.1cm

For simple compact simply-connected groups $\,G$, \,the multiplicative 
structure on $\,\CG_k\,$ always exists and 
is unique up to isomorphism \cite{WMult}. In the present 
paper, we address the question of obstructions to the existence 
of a multiplicative structure on the gerbe $\,\CG_k\,$ over simple compact 
non-simply connected groups $\,G\,$ with fundamental group $\,\pi_1(G)=Z$, 
\,as well as the classification of such structures. 

We show that the unique obstruction is provided by the $\,U(1)$-valued phases
$\,c_{\varphi_1,\varphi_2}\,$ that appear in the formula
\qq
{\CH}_{\CG_k}(\varphi_1\varphi_2)\ =\ c_{\varphi_1,\varphi_2}\,\,
{\CH}_{\CG_k}(\varphi_1)\,\,{\CH}_{\CG_k}(\varphi_2)\,\,
\ee^{2\pi i\int\limits_\Sigma(\varphi_1\times 
\varphi_2)^*\omega_k}
\label{PW0}
\qqq
that relates the holonomy of the point-wise product
of two group-valued fields $\ \varphi_{1,2}:\Sigma\rightarrow G\ $ to the 
product of the individual holonomies. Above,  
\qq
\omega_k\ =\ \frac{k}{8\pi^2}\tr\,(g_1^{-1}dg_1)(g_2dg_2^{-1})
\label{rho}
\qqq
is a 2-form on the double group $\,G\times G\equiv G^2$.
\,That Eq.\,(\ref{PW0}) holds with $\,c_{\varphi_1,\varphi_2}\equiv1\,$ for 
simply connected groups $\,G\,$ is the content of the 
Polyakov-Wiegmann formula that for the first time appeared (in an equivalent 
form) in \cite{PW}, see also \cite{FSh}. Its generalization to the 
non-simply-connected groups $\,G\,$ was obtained in \cite{FGK} 
where the phases $\,c_{\varphi_1,\varphi_2}\,$ were computed for the surface 
$\,\Sigma\,$ of genus $1$. \,In the latter case, they reduce to a certain 
$\,U(1)$-valued 2-cocycle $\,c\,$ on the group $\,Z\times Z\equiv Z^2\,$ 
that we shall call, accordingly, the FGK cocycle.
\,Our main result states that a multiplicative structure on the gerbe 
$\,\CG_k\,$ over a non-simply-connected group $\,G\,$ exists if and only 
if the FGK cocycle is identically equal to 1. \,Under this condition,
such a structure on $\,\CG_k\,$ is unique up to isomorphism. 
Thus, the computation of the FGK cocycle for all simple compact Lie groups 
in \cite{FGK} provides a complete classification of multiplicative structures 
on the gerbes $\mathcal{G}_k$. 

\vskip 0.1cm

The paper is organized as follows. In Sec.\,\ref{sec:cohob}, we recall that
a multiplicative structure on the gerbe $\,\CG_k\,$ requires,
in particular, that a certain gerbe with vanishing curvature (i.e. flat) 
over the group $\,G^2$, \,constructed from the gerbe 
$\,\CG_k\,$ over $\,G$, \,be trivial. \,Isomorphism classes of flat 
gerbes over $\,G^2\,$ may be identified \cite{top} with cohomology 
classes of $\,U(1)\,$-valued 2-cocycles on the group $\,G^2$.
\,Consequently, such classes provide cohomological obstructions to 
the triviality of flat gerbes. The corresponding 
cohomology group is calculated by standard tools of homological algebra. 
\,On the other hand, a flat gerbe over $\,G^2\,$ is trivial if and only if 
its holonomy is trivial. For the flat gerbe mentioned above, the latter 
property is equivalent to the strict Polyakov-Wiegmann formula without 
additional phases that may appear in the general case (\ref{PW0}). We 
explain in Sec.\,\ref{sec:PW} how such phases give 
rise to the FGK 2-cocycle. In Sec.\,\ref{sec:calFGK}, we recall from 
\cite{FGK} the 
calculation of this cocycle and in Sec.\,\ref{sec:cohFGK}, we clarify
the relation between the cohomological obstruction classes  and
the FGK cocycles by connecting both to bihomomorphisms 
in $\,\mathrm{Hom}(Z\otimes 
Z,U(1))$. \,Such bihomomorphisms appeared in the algebraic approach 
\cite{KS,FRS} to simple current orbifolds of the WZW models.
\vskip 0.1cm

The following sections of the paper are devoted to a more thorough discussion 
of multiplicative gerbes. 
In Sec.\,\ref{sec:emgerbes}, after some preparations, 
we formulate an abstract definition
of a multiplicative gerbe equivariant with respect to the action of a discrete 
group. This is done in a way that allows to view multiplicative gerbes 
over non-simply-connected groups $\,G\,$ as multiplicative gerbes over their 
universal covers $\,\tilde G\,$ that are equivariant under the deck action 
of $\,Z=\pi_1(G)$. 
Sec.\,\ref{sec:local} describes  
equivariant multiplicative gerbes in terms of local data. 
The local description 
permits an analysis of obstructions to the existence of equivariant
multiplicative gerbes that we perform in Sec.\,\ref{sec:obs}. 
We show that in the case of multiplicative gerbes over the group 
$\,\tilde G\,$ equivariant under the deck action of $\,Z$, \,the
only obstructions that may be non-trivial belong to the cohomology groups 
$\,H^3(Z,U(1))\,$ and $\,H^2(Z^2,U(1))$. \,The first one obstructs the 
existence of the gerbe $\,\CG_k\,$ over the group $\,G=\tilde G/Z\,$ and was 
studied in detail in \cite{GR04}. \,The second one in $\,H^2(Z^2,U(1))\,$
is the cohomological obstruction, mentioned above, to the existence of 
a multiplicative structure on the gerbe $\,\CG_k$. \,Its triviality is 
equivalent to the triviality of the FGK 2-cocycle. Finally, in 
Sec.\,\ref{sec:unique}, we discuss  equivalences of equivariant 
multiplicative gerbes and prove that all multiplicative structures on the 
fixed gerbe $\,\CG_k\,$ are isomorphic. Conclusions summarize the
results of the paper and discuss perspectives for the further work.   

\medskip

%KG: I have rewritten the acknowledgements:
\paragraph*{Acknowledgements.} \,In early stages of the collaboration,
the authors profited from discussions with Rafa\l \ Suszek. K.G.'s work was 
a part of the project ANR-05-BLAN-0029-03. K.W. thanks the ANR for support 
during visits at the ENS-Lyon. He gratefully acknowledges a Feodor Lynen 
scholarship granted by the Alexander von Humboldt foundation. 

\section{Cohomological Obstructions}
\label{sec:cohob}

\noindent The principal geometric objects that we shall deal with 
in this paper are hermitian bundle gerbes with  unitary connection over 
a manifold $\,M\,$ \cite{Murr,MurrS}, called below ``gerbes'' for 
short. The curvature of a gerbe is a closed 3-form over $\,M$. 
\,Gerbes over $\,M\,$ form a 2-category \cite{Ste00} with objects, 
1-morphisms  between objects (called also ``stable morphisms'' or 
simply ``morphisms'') and 2-morphisms between 1-morphisms, 
see also Sec.\,2.1 of \cite{GSW}. \,One may define tensor product 
of gerbes, their duals and their pullbacks. 
The isomorphism classes of flat gerbes form a group that may be 
naturally identified with the cohomology group $\,H^2(M,U(1))\,$  
\cite{top,GR02}. \,We shall study gerbes, equipped with additional 
structures, over Lie groups.
\vskip 0.1cm
 
Let $\,\tilde G\,$ be a simple, compact, connected and simply-connected 
Lie group and let $\,Z\,$ be a subgroup of its center: $\,Z\subset 
Z(\tilde G)$. \,The possible cases are $\,Z=\NZ_N\,$ for some $\,N\geq1\,$ or 
$\,Z=\NZ_2^2$. \,The non-cyclic case occurs for 
$\,Z=Z(Spin(4r))$. \,More complicated discrete Abelian groups appear 
if one admits non-simple groups $\,\tilde G\,$ that will not be discussed 
here. We shall consider the quotient Lie groups $\,G=\tilde G/Z\,$ that are
non-simply connected for non-trivial subgroups $\,Z\,$ since $\,\pi_1(G)=Z$.
The deck action of $\,Z\,$ on $\,\tilde G\,$ may be identified with its action
by the group multiplication. \,Eq.\,(\ref{meg}) defines closed bi-invariant 
3-forms $\,H_k\,$ on $\,G\,$ that pull back to  3-forms $\,\tilde H_k\,$ 
on $\,\tilde G\,$ given by the same formula. 
\vskip 0.1cm

Let $\,\CG_k\,$ be a gerbe with  curvature $\,H_k\,$ over 
$\,G$. \,Such a gerbe exists if and only if the 3-form $\,H_k\,$
is integral (i.e. has integral 3-periods). The normalization in
Eq.\,(\ref{meg}) is chosen so that this happens for $\,k\in\NZ\,$ if
the subgroup $\,Z\,$ is trivial but only for certain levels 
$\,k\in\NZ\,$ for non-trivial $\,Z$. \,Gerbes $\,\CG_k$, \,when they exist, 
are unique up to  isomorphism except for $\,G=Spin(4r)/\NZ_2^2
=SO(4r)/\NZ_2\,$ where, for each $\,k\in2\NZ\,$ if $\,r\,$ is 
odd and for each $\,k\in\NZ\,$ if $\,r\,$ is even, there are two 
isomorphism classes of gerbes with curvature $\,H_k$ \cite{GR04}. 
\,As already mentioned in the Introduction, gerbes $\,\CG_k\,$ are employed 
in the definition of Feynman amplitudes in the Wess-Zumino-Witten (WZW) 
sigma models with target groups $\,G$, \,see also \cite{top,GR02,G05}.
\vskip 0.2cm

We shall use the notion of a multiplicative gerbe over Lie groups
in the version that appeared in \cite{WMult} as a refinement 
of the concept introduced in \cite{CJMSW}. A multiplicative gerbe 
may be viewed as an ordinary gerbe over a Lie group equipped with 
a multiplicative structure. The latter assures the compatibility 
of the gerbe with the group multiplication. 
As explained in \cite{CJMSW} and \cite{WMult}, the gerbe 
$\,\CG_k\,$ over group $\,G\,$ equipped with a multiplicative structure 
canonically determines Feynman amplitudes in the CS theory with gauge 
group $\,G$. \,A precise definition of a multiplicative gerbe may be found
         in Sec.\,\ref{sec:emgerbes} below. Here, we shall only briefly
         elucidate this notion. First, let us remark that the curvature
         form $\,H_k\,$ of the gerbe $\,\CG_k\,$ satisfies the identity
         \qq
         m^*H_k\ =\ p_1^*H_k\,+\,p_2^*H_k\,+\,d\omega_k\,,
         \label{3H}
         \qqq
         where $\,m:G^2\rightarrow G\,$ is the group multiplication,
         $\,p_{1,2}:G^2\rightarrow G\,$ are the projections and the 2-form
         $\,\omega_k\,$ on $\,G^2\,$ is given by Eq.\,(\ref{rho}).
         Explicitly, the relation (\ref{3H}) boils down to equality
         \qq
         \tr\,((g_1g_2)^{-1}d(g_1g_2))^3\,=\,\tr\,(g_1^{-1}dg_1)^3\,+\,
         \tr\,(g_2^{-1}dg_2)^3\,+\,3\,d\,\tr\,(g_1^{-1}dg_1)(g_2dg_2^{-1})
         \qqq
         that is straightforward to check. A multiplicative structure
         over the gerbe $\,\CG_k\,$ realizes a lift of the relation
         (\ref{3H}) from the level of curvature 3-forms to the level
         of gerbes. More precisely, it involves an isomorphism
         \qq
         m^*\CG_k\ \,\cong\,\ p_1^*\CG_k\otimes p_2^*\CG_k\otimes
         \CI_{\omega_k}
         \label{2grbs}
         \qqq
         between  gerbes over $\,G^2$, \,where $\,\CI_{\omega_k}\,$ is
         a gerbe over $\,G^2\,$ which is trivial except for the global
         curving 2-form $\omega_k$ and the corresponding curvature 3-form
         $d\omega_k$. \,The remaining part of the definition of a
         multiplicative structure consists of certain associativity data
         for the isomorphism (\ref{2grbs}).
\vskip 0.2cm

The existence of an isomorphism (\ref{2grbs}) is equivalent to the 
statement that the isomorphism class $\,\kappa\,$ of the flat gerbe 
\qq
 m^*\CG_k\otimes p_1^*\CG_k^*\otimes p_2^*\CG_k^*\otimes
\CI_{-\omega_k}\ \ \equiv
\ \ \CK_k
\label{flatg}
\qqq
on $\,G^2\,$ is trivial. \,Above, $\,\CG_k^*\,$ denotes 
the gerbe dual to $\,\CG_k\,$ with the inverse holonomy and opposite 
curvature. \,The isomorphism classes of flat gerbes on $\,G^2\,$ form the 
cohomology group $\,H^2(G^2,U(1))\,$ so that the isomorphism 
(\ref{2grbs}) exists if and only if the class $\,\kappa\in H^2(G^2,U(1))\,$ 
associated to $\mathcal{K}_k$
vanishes. In other words, $\,\kappa\,$ is the cohomological obstruction 
to the existence of the isomorphism (\ref{2grbs}).
\vskip 0.1cm

We have to understand the structure of the obstruction cohomology group
$\,H^2(G^2,U(1))\,$ for the Lie groups $\,G=\tilde G/Z$. 
\,The lowest homology groups (with integer coefficients) for simple, compact, 
connected and simply-connected Lie group $\,\tilde G\,$ are
\qq
H_0(\tilde G)\,=\,\NZ\,,\qquad H_1(\tilde G)\,=\,0\,=\,H_2(\tilde G)\,.  
\qqq
Since $\,Z\,$ acts properly on $\,\tilde G$, \,it follows 
(see e.g. Corollary 7.3 of \cite{McL}) that 
\qq
H_n(G) =\ H_n(Z)\qquad{\rm for}\qquad n=0,1,2\,, 
\label{HnG'}
\qqq
where on the right hand side appear the group-homology groups 
\cite{McL,Brown}. \,One has
\qq
H_0(Z)\,=\,\NZ\,,\qquad H_1(Z)\,=\,Z\,,\qquad H_2(Z)\,=\,
\begin{cases}
\hbox to 1.5cm{\,$0$\hfill}{\rm if}\qquad Z=\NZ_N\,,\cr\cr
\hbox to 1.5cm{\,$\NZ_2$\hfill}{\rm if}\qquad Z=\NZ_2^2\,.
\end{cases} 
\label{HnZ}
\qqq
The Universal Coefficients Theorem implies that
\qq
H^2(G,U(1))\ \cong\ \mathrm{Hom}(H_2(G),U(1))
\ \cong\ \mathrm{Hom}(H_2(Z),U(1))\ \cong\ H^2(Z,U(1))\,,
\label{h2G'}
\qqq
where, in the last member, $\,U(1)\,$ is considered as a trivial
$\,Z$-module. This gives the result
\qq
H^2(G,U(1))\ \cong\ \begin{cases}\hbox to 1.5cm{\,$0$\hfill}{\rm if}
\qquad Z=\NZ_N\,,\cr\cr
\hbox to 1.5cm{\,$\NZ_2$\hfill}{\rm if}\qquad Z=\NZ_2^2
\end{cases}
\label{ucth0}
\qqq
discussed in detail in \cite{FGK}.
\vskip 0.2cm

Low homology and cohomology of the product groups 
$\,\tilde G^2\,$ and $\,G^2\,$ can be computed similarly. One has:
\qq
H_0(\tilde G^2)\,=\,\NZ\,,\qquad H_1(\tilde G^2)\,=\,0\,=\,H_2(\tilde G^2)\,.
\qqq
and 
\qq
H_n(G^2)\ \cong\ H_n(Z^2)\qquad{\rm for}\qquad n=0,1,2\,,
\qqq
and, from the Universal Coefficients theorem,
\qq
H^2(G^2,U(1))\ \cong\ H^2(Z^2,U(1))\,,
\label{ucth}
\qqq
where $\,U(1)\,$ is considered as a trivial
$\,Z^2$-module. \,The groups $\,H_n(G^2)\cong H_n(Z^2)\,$ for $\,n\leq2\,$ 
are easy to compute from the K\"unneth formula: 
\qq
H_0(G^2)\ &\cong&\ \NZ\,,\cr\cr
H_1(G^2)\ &\cong&\ H_1(G)\oplus H_1(G)
\ \,\cong\ \,Z^2\,,\label{H11}\\
\cr
H_2(G^2)\ &\cong&\ H_2(G)\oplus H_1(G)\otimes H_1(G)\oplus
H_2(G)\label{H222}\,.
\qqq
More precisely, in the last isomorphism, the injection of the two 
components $\,H_2(G)\,$ 
into $\,H_2(G^2)\,$ is induced by the embeddings $\,G\ni g\mapsto 
(g,1)\in G^2\,$ and $\,G\ni g\mapsto(1,g)\in G^2\,$  
whereas the injection of $\,H_1(G)\otimes H_1(G)\,$
is given by the cross product. 
\,Eqs.\,(\ref{HnG'}), (\ref{HnZ}) and (\ref{H222})
give then the result:
\qq
H_2(G^2)\ \cong\ \begin{cases}\hbox to 1.5cm{\,$\NZ_N$\hfill}{\rm if}
\qquad Z=\NZ_N\,,\cr\cr
\hbox to 1.5cm{\,$\NZ_2^6$\hfill}{\rm if}\qquad Z=\NZ_2^2\,.
\end{cases}
\label{H22} 
\qqq
The Universal Coefficients Theorem implies now that 
\qq
H^2(G^2,U(1))\ &\cong&\ \mathrm{Hom}(H_2(G^2),U(1))\cr\cr
&\cong&\ \mathrm{Hom}(H_2(G)\oplus H_1(G)\otimes H_1(G)\oplus H_2(G)\,,
\,U(1))\cr\cr
&\cong&\ H^2(G,U(1))\oplus \mathrm{Hom}(H_1(G)\otimes H_1(G),
U(1))\oplus H^2(G,U(1))\cr\cr
&\cong&\ \begin{cases}\hbox to 1.5cm{\,$\NZ_N$\hfill}{\rm if}
\qquad Z=\NZ_N\,,\cr\cr
\hbox to 1.5cm{\,$\NZ_2^6$\hfill}{\rm if}\qquad Z=\NZ_2^2\,.
\end{cases}
\label{H22U}
\qqq
Eqs.\,(\ref{ucth}), (\ref{HnG'}), (\ref{HnZ}) and (\ref{h2G'}) permit
to rewrite the latter isomorphisms in terms of the cohomology of 
finite groups:
\qq
H^2(Z^2,U(1))\ \cong\ H^2(Z,U(1))\oplus
\mathrm{Hom}(Z\otimes Z,U(1))\oplus H^2(Z,U(1))\,.
\qqq
Here, the injections of $\,H^2(Z,U(1))\,$ into 
$\,H^2(Z^2,U(1))\,$ are induced by considering the
$\,U(1)$-valued 2-cocycles $\,c^Z_{z,z'}\,$ on $\,Z\,$ as 2-cocycles
on $\,Z^2\,$ according to the formulae
\qq
c^{Z^2}_{(z_1,z_2),(z'_1,z'_2)}\,=\,c^Z_{z_1,z'_1}\qquad
{\rm or}\qquad c^{Z^2}_{(z_1,z_2),(z'_1,z'_2)}\,=\,c^Z_{z_2,z'_2}\,,
\label{cZZ}
\qqq
respectively, whereas the injection of the group 
$\,\mathrm{Hom}(Z\otimes Z,U(1))\,$ 
of bihomomorphisms $\,\zeta:Z^2\rightarrow U(1)\,$
into the cohomology group $\,H^2(Z^2,U(1))\,$ may be induced 
by setting
\qq
c^{Z^2}_{(z_1,z_2),(z'_1,z'_2)}\,=\,\zeta(z_1,z'_2)\,.
\label{zZZ}
\qqq
The above information about the cohomology group $\,H^2(Z^2,U(1))
\cong H^2(G^2,U(1))\,$ will be used in the sequel.

\section{Generalized Polyakov-Wiegmann Formula and the FGK Cocycle}
\label{sec:PW} 

\no Isomorphic gerbes have the same holonomy. The converse is also
true 
over manifolds for which $\,H^2(M,U(1))=\mathrm{Hom}(H_2(M),U(1))\,$ as
is the case for groups $\,G\,$ or $\,G^2\,$ (this uses the fact that
$\,H_2(M)\,$ is spanned by images of closed oriented surfaces). 
The triviality up to isomorphism of the gerbe $\,\CK_k\,$ of 
Eq.\,(\ref{flatg}) is then equivalent to the triviality of its holonomy
\qq
\CH_{\CK_k}(\varphi_1\times\varphi_2)\ =\ 
\CH_{\CG_k}(\varphi_1\varphi_2)\ \CH_{\CG_k}(\varphi_1)^{-1}\ 
\CH_{\CG_k}(\varphi_2)^{-1}\ \ee^{-2\pi i\int\limits_\Sigma(\varphi_1\times 
\varphi_2)^*\omega_k}\ =\ 
c_{\varphi_1,\varphi_2}\ \in\,U(1)
\label{cgg}
\qqq
for any pair of maps $\,\varphi_{1,2}:\Sigma\rightarrow G$, \,see 
Eq.\,(\ref{PW0}). \,Somewhat surprisingly, 
the generalized Polyakov-Wiegmann formula (\ref{PW0}) leads to a 
different picture of obstructions to the existence of an isomorphism 
(\ref{2grbs}) than the discussion in Sec.\,\ref{sec:cohob}. Namely, 
it induces obstructions living in the group of $\,U(1)$-valued 2-cocycles 
on $\,Z^2\,$ and not in the corresponding cohomology group 
$\,H^2(Z^2,U(1))\,\cong\,H^2(G^2,U(1))$. \,Such obstruction 2-cocycles 
may be nontrivial even if their cohomology class is trivial. \,Here is how 
this story goes. 
\vskip 0.2cm

\no{\bf1}.\ \ First, if $\,\varphi_{1,2,3}:\Sigma\rightarrow G\,$ then
the holonomy of $\,\CH_{\CG_k}(\varphi_1\varphi_2\varphi_3)\,$ may be
calculated in two different ways. On the one hand, 
\qq
\CH_{\CG_k}(\varphi_1\varphi_2\varphi_3)&=&c_{\varphi_1\varphi_2,\varphi_3}\ 
\CH_{\CG_k}(\varphi_1\varphi_2)\ \CH_{\CG_k}(\varphi_3)\ 
\ee^{\,2\pi i\int\limits_\Sigma(\varphi_1\varphi_2\times 
\varphi_3)^*\omega_k}\cr\cr
&=&c_{\varphi_1,\varphi_2}\ c_{\varphi_1\varphi_2,\varphi_3}\ 
\CH_{\CG_k}(\varphi_1)\ \CH_{\CG_k}(\varphi_2)\ \CH_{\CG_k}(\varphi_3)\cr\cr 
&&\quad\cdot\ \ee^{\,2\pi i\int\limits_\Sigma[(\varphi_1\varphi_2\times 
\varphi_3)^*\omega_k\,+\,(\varphi_1\times\varphi_2)^*\omega_k]}.
\label{1}
\qqq
On the other hand,
\qq
\CH_{\CG_k}(\varphi_1\varphi_2\varphi_3)&=&c_{\varphi_1,\varphi_2\varphi_3}\ 
\CH_{\CG_k}(\varphi_1)\ \CH_{\CG_k}(\varphi_2\varphi_3)\ 
\ee^{\,2\pi i\int\limits_\Sigma(\varphi_1\times\varphi_2\varphi_3)^*
\omega_k}\cr\cr
&=&c_{\varphi_1,\varphi_2\varphi_3}\ c_{\varphi_2,\varphi_3}\ 
\CH_{\CG_k}(\varphi_1)\ \CH_{\CG_k}(\varphi_2)\ \CH_{\CG_k}(\varphi_3)\ \cr\cr
&&\quad\cdot\ \ee^{\,2\pi i\int\limits_\Sigma[(\varphi_1\times 
\varphi_2\varphi_3)^*\omega_k\,
+\,(\varphi_2\times \varphi_3)^*\omega_k]}.
\label{2}
\qqq
Since a direct calculation shows that on $\,G^3$,
\qq
(m\circ p_{12}\times Id)^*\omega_k\,+\,p_{12}^*\omega_k\ =\ 
(Id\times m\circ p_{23})^*\omega_k\,+\,p_{23}^*\omega_k
\label{omegadeltaclosed}
\qqq
with the natural notation for the projections $\,p_{ij}:G^3\rightarrow 
G^2$, \,we infer that the exponential terms on 
the right hand side of Eqs.\,(\ref{1}) and (\ref{2}) coincide and, 
consequently, that 
\qq
c_{\varphi_1,\varphi_2}\ c_{\varphi_1\varphi_2,\varphi_3}\ =\ c_{\varphi_1,
\varphi_2\varphi_3}\ c_{\varphi_2,\varphi_3}\,,
\qqq 
i.e. that $\,c_{\varphi_1,\varphi_2}\,$ is a $U(1)$-valued 2-cocycle on the
group of maps $\,\varphi:\Sigma\rightarrow G$. 
\vskip 0.1cm

\begin{figure}[ht]
\begin{center}
\vskip -0.1cm 
\leavevmode 
\epsfxsize=8cm %% {where n is the figure width you want} 
\epsfysize=2.5cm %% {where n is the figure height you want} 
\epsfbox{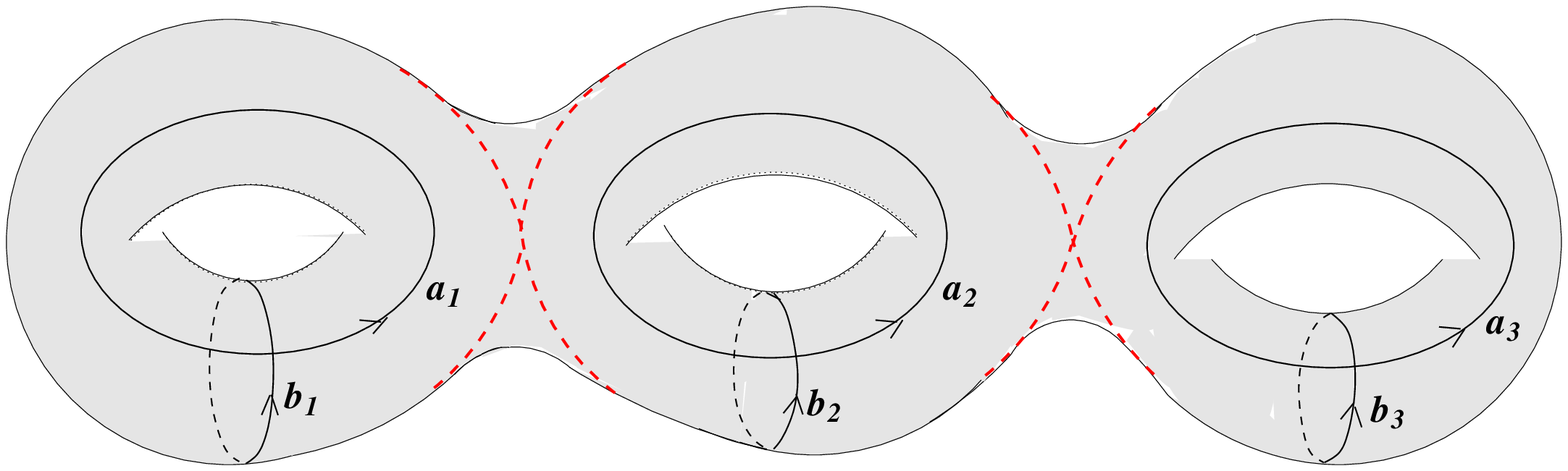}%\hspace*{0.5cm} 
\end{center} 
\caption{Genus 3 surface with a marking; broken red lines indicate the 
contours of its version with pinched off handles}
\label{fig:fig1} 
\end{figure}

\no{\bf2}.\ \ Second, $\,c_{\varphi_1,\varphi_2}\,$ depends only on the
homotopy classes of $\,\varphi_1\,$ and $\,\varphi_2$. \,Indeed, if
$\,\varphi_1\,$ is homotopic to $\,\varphi'_1\,$ and $\,\varphi_2\,$ 
is homotopic to $\,\varphi'_2$, \,with 
$\,\phi_{1,2}:[0,1]\times\Sigma\rightarrow G\,$ being the corresponding
homotopies, \,then by Eq.\,(\ref{holcurv}), 
\qq
&&\CH_{\CG_k}(\varphi'_{1})\ =\ \CH_{\CG_k}(\varphi_1)\ \ee^{\,2\pi i
\int\limits_{[0,1]\times\Sigma}\phi_1^*H_k},\cr
&&\CH_{\CG_k}(\varphi'_{2})\ =\ \CH_{\CG_k}(\varphi_2)\ \ee^{\,2\pi i
\int\limits_{[0,1]\times\Sigma}\phi_2^*H_k},\cr
&&\CH_{\CG_k}(\varphi'_{1}\varphi'_{2})\ =\ \CH_{\CG_k}(\varphi_1\varphi_2)\ 
\ee^{\,2\pi i\int\limits_{[0,1]\times\Sigma}(\phi_1\phi_2)^*H_k}.
\nonumber
\qqq
Using the relation (\ref{3H}),\ we infer that
\qq
c_{\varphi'_1,\varphi'_2}\ =\ c_{\varphi_1,\varphi_2}\ \equiv\ 
c_{[\varphi_1],[\varphi_2]}\,,
\nonumber
\qqq
where $\,[\varphi]\,$ denotes the homotopy class of 
the map $\,\varphi:\Sigma\rightarrow G$.
\,Such homotopy classes are in one-to-one correspondence with elements
of $\,Z^{2\gamma}$, \,where $\,\gamma\,$ is the genus of $\Sigma$.
\,The element $\,(z_1,z_2,...,z_{2\gamma-1},z_{2\gamma})\,$ corresponding
to $\,[\varphi]\,$ is given by the holonomies
\qq
z_{2i-1}\,=\,\CP\,\ee^{\,\int\limits_{a_i}A_\varphi},\qquad
z_{2i}\,=\,\CP\,\ee^{\,\int\limits_{b_i}A_\varphi},
\label{hols}
\qqq
of the non-Abelian flat gauge field $\,A_\varphi=\varphi^*(g^{-1}dg)\,$
on $\,\Sigma$. \,Above, $\,\CP\,$ stands for the path-ordering 
(from left to right) along paths 
$\,a_i,b_i$, $\,i=1,\dots,\gamma$, \,that generate a fixed marking 
of $\,\Sigma$, \,see Fig.\,\ref{fig:fig1}. 
Note that 
$\,[\varphi_1\varphi_2]=[\varphi_1][\varphi_2]$, 
\,where on the right hand side the product is taken in $\,Z^{2\gamma}$.
\,We infer that $\,c_{[\varphi_1],[\varphi_2]}\,$ is a $U(1)$-valued 2-cocycle 
on the finite group $\,Z^{2\gamma}$. 
\vskip 0.2cm

\no{\bf3}.\ \ Third, it is easy to see from the definition of the 2-cocycle 
$\,c_{[\varphi_1],[\varphi_2]}\,$ that if $\,[\varphi_1]=(z_1,\dots,
z_{2\gamma})\,$ and $\,[\varphi_2]=(z'_1,\dots,z'_{2\gamma})\,$ then
\qq
c_{(z_1,\dots,z_{2\gamma}),(z'_1,\dots,z'_{2\gamma})}\ =\ 
\prod\limits_{i=1}^\gamma c_{(z_{2i-1},z_{2i}),(z'_{2i-1},z'_{2i})}\,.
\label{red}
\qqq
In order to obtain this relation, just calculate 
$\,c_{[\varphi_1],[\varphi_2]}\,$ from the holonomy of the fields defined 
on a family of surfaces $\,\Sigma\,$ whose handles are pinched away from 
each other, see Fig.\,\ref{fig:fig1}.
On the one hand, $\,c_{[\varphi_1],[\varphi_2]}\,$ is the same
for all surfaces in the family because the holonomies (\ref{hols}) are
the same. On the other hand, since the fundamental 
group of $\,G\,$ is commutative, the holonomy of 
$\,A_\varphi\,$ around the pinched curves is trivial and 
one may take fields that extend smoothly to the limiting surface with 
pinched handles giving rise the the product expression on the right 
hand side. It is then enough to consider surface $\,\Sigma\,$
of genus 1, i.e. the torus $\,\NT^2=S^1\times S^1$, \,leading to a 
$\,U(1)$-valued 2-cocycle $\,c_{(z_1,z_2),(z'_1,z'_2)}\,$ on the 
finite group $\,Z^2$, the FGK cocycle.
\,Let us stress that it is the non-triviality of the 
FGK cocycle $\,c_{(z_1,z_2),(z'_1,z'_2)}\,$ \,and \emph{not} of 
its cohomology class in $\,H^2(Z^2,U(1))$, \,that obstructs the existence 
of a multiplicative structure on the gerbe $\,\CG_k\,$ over the group 
$\,G=\tilde G/Z$. 

\vskip0.2cm

It is well known that 2-cocycles on a group are related to projective
representations. In particular, the FGK 2-cocycle $\,c\,$ is related
to a projective representation $\,\,\Psi\mapsto{}^{(z_1,z_2)}\Psi\,\,$
of $\,Z^2\,$ in the space of quantum states of the group 
$\,\tilde G\hspace{0.075cm}$ CS theory on the 3-manifold $\,\NT^2\times\NR\,$
\cite{G90}. \,This space 
is spanned by the characters of the central extension of the 
loop group $\,L\tilde G$. \,The FGK 2-cocycle 
characterizes the projectivity of the representation:
\qq
{}^{(z_1z_1',z_2z_2')}\Psi\ =\ c_{(z_1,z_2),(z_1',z_2')}\,{}^{(z_1,z_2)}
({}^{(z'_1,z'_2)}\Psi)\,.
\nonumber
\qqq
If $\,c\equiv 1\,$ then the representation of $\,Z^2\,$ is genuine rather
than projective. In this case, one may define the subspace of the 
$\,Z^2$-invariant states which forms the space of quantum 
states of the CS theory with the non-simply-connected group 
$\,G=\tilde G/Z\,$ on the same manifold $\,\NT^2\times\NR$. \,The subspace 
of the $\,Z^2$-invariant 
states  $\,\Psi\,$ is spanned by the characters of the central extension 
of the loop group $\,LG\,$ that is determined by the corresponding 
multiplicative gerbe. As already mentioned, such a central extension provides 
the extended chiral algebra of the corresponding group $\,G\,$ WZW model. 
\,In particular, when $\,c\equiv 1$, \,the toroidal partition function 
of the group $\,G\,$ WZW model is a diagonal combination of the absolute 
values squared of the characters of the extended chiral algebra.

\section{Calculation of the FGK Cocycle} 
\label{sec:calFGK}

\no The calculation of the 2-cocycles $\,c_{(z_1,z_2),(z'_1,z'_2)}\,$
to which case Eq.\,(\ref{red}) reduces the general expression, has been 
done in ref. \cite{FGK}. Let us recall (and slightly complete) the argument 
of \cite{FGK}.
\vskip 0.2cm

We shall start from the cyclic case when $\,Z=\NZ_N\,$ with the 
generator
$\,\zeta=\ee^{2\pi i\,\theta}\,$ for some $\,\theta\not=0\,$ in the coweight 
lattice $\,P^\vee\,$ of group $\,G$. \,Recall that 
the existence of a gerbe 
$\,\CG_{k}\,$ with curvature $\,H_k\,$ over the group 
$\,G=\tilde G/Z\,$ requires the integrality of $\,H_k$. \,As was shown 
in \cite{FGK}, the latter is equivalent to the condition
\qq
\frac{_1}{^2}kN\,\tr\,\theta^2\ \in\ \NZ
\label{condN}
\qqq  
that selects the admissible levels $\,k\in\NZ$. 
\,To each pair $\,(m,n)\,$ of integers, we may associate a field 
configuration $\,\varphi_{m,n}:\NT^2\rightarrow G\,$ on the 
two-dimensional torus $\,\NT^2\,$ such that
\qq
\varphi_{m,n}(\ee^{i\sigma_1},\ee^{i\sigma_2})\ =\ 
\ee^{i(\sigma_1m\theta+\sigma_2n\theta)}.
\label{27}
\qqq
Note that the homotopy class of $\,\varphi_{m,n}\,$
\qq
[\varphi_{m,n}]\ =\ (\zeta^m,\zeta^n)\,\in\,Z^2\,.
\label{gmn}
\qqq
Since $\,\varphi_{m,n}\,$ takes values in the circle $\,\{\,\ee^{i\sigma\theta}
\in G\,\,|\,\,\sigma\in[0,2\pi[\,\}\,$ and all gerbes over $\,S^1\,$ 
are trivial up to isomorphism, it follows that
\qq
\CH_{\CG_k}(\varphi_{m,n})\ =\ 1\,.
\label{mntr}
\qqq
As a result,
\qq
c_{(\zeta^m,\zeta^n),(\zeta^{m'},\zeta^{n'})}\ &=& 
\ \ee^{-\frac{ik}{4\pi}\int\limits_{\NT^2}\tr\,(\varphi_{mn}^{-1}d\varphi_{mn})
(\varphi_{m'n'}d\varphi^{-1}_{m'n'})}\ =\ \ee^{-\frac{ik}{4\pi}
\int\limits_0^{2\pi}\int\limits_0^{2\pi}
\tr\,(m\theta d\sigma_1+n\theta d\sigma_2)(m'\theta d\sigma_1
+n'\theta d\sigma_2)}\cr\cr
&=&\ \ee^{\pi i\,k\,(m'n-mn')\,
\tr\,\theta^2}\ \,\equiv\,\ c_{(m,n),(m',n')}\,.
\label{czz}
\qqq
Note that the relation (\ref{condN}) implies directly 
that $\,c_{(m,n),(m',n')}\,$ depends only on the classes modulo $\,N\,$ 
of the integers $\,m,n,m',n'$. 
\,Besides, $\,c_{(m,n),(m',n')}\equiv 1\,$
if and only if
\qq
\frac{_1}{^2}k\,\tr\,\theta^2\ \in\ \NZ\,.
%\label{cond1}
\nonumber
\qqq  
Let us observe that in the particular case when $\,N=2\,$ 
and $\,k\,\tr\theta^2\,$ is an integer, one has:
\qq
c_{(m,n),(m',n')}\ =\ e_{(m,n)}\ e^{-1}_{(m+n,m'+n')}\ e_{(m',n')}
\nonumber
\qqq
for $\,e_{(m,n)}=\ee^{\,\pi i\,k\,mn\,\tr\,\theta^2}\,$ so that the 
FGK 2-cocycle is cohomologically trivial although it is non-trivial 
if $\,k\,tr\,\theta^2\,$ is an odd integer as for $\,Z=Z(SU(2))\,$
at even levels not divisible by $\,4$. 
\,This shows that the requirement
of triviality of the FGK cocycle is, in general, strictly stronger than
the requirement of its cohomological triviality. 
\vskip 0.2cm

Consider now the case when $\,Z=Z(Spin(4r))=\NZ_2^2\,$ and 
is generated by $\,\zeta_1=\ee^{2\pi i\,\theta_1}\,$ and 
$\,\zeta_2=\ee^{2\pi i\,\theta_2}\,$
for certain $\,\theta_1,\,\theta_2\in P^\vee$. \,As was shown 
in \cite{FGK}, the integrality of the 3-form $\,H_k\,$ on $\,G
=Spin(4r)/\NZ_2^2\,$ imposes now the conditions
\qq
k\,\tr\,\theta_1^2,\,\ k\,\tr\,\theta_2^2,\,\ 2k\,
\tr\,\theta_1\theta_2\,\in\,\NZ
\label{cond'}
\qqq
(ref.\,\cite{FGK} considered an additional restriction 
that required that $\,k\,\tr\,\theta_1\theta_2\,$ be integral; we 
drop it here). An inspection shows that 
$\,\tr\,\theta_1\theta_2\,$ is always a half-integer and that 
$\,\tr\,\theta_1^2\,$ and $\,\tr\,\theta_2^2\,$ are integers
when $\,r\,$ is even and, say, the first one is a half-integer and 
the second one an integer when $\,r\,$ is odd. \,It follows that 
the gerbes $\,\CG_k\,$ over $\,Spin(4r)/\NZ_2^2\,$
exist for all $\,k\in\NZ\,$ if $\,r\,$ is even and for all
$\,k\in2\NZ\,$ if $\,r\,$ is odd, as already indicated in 
Sec.\,\ref{sec:cohob}. \,Eq.\,(\ref{gmn}) with $\,m\theta\,$ 
standing now for 
$\,m_1\theta_1+m_2\theta_2\,$ and $\,n\theta\,$ for $\,n_1\theta_1
+n_2\theta_2\,$ associates a field configuration
$\,\varphi_{m,n}:\NT^2\rightarrow G\,$ to a pair $\,(m,n)\,$ of vectors 
in $\,\NZ^2\,$ with $\,m=(m_1,m_2)\,$ and $\,n=(n_1,n_2)$.
\,The relation (\ref{gmn}) still hold with 
$\,\zeta^m\equiv\zeta_1^{m_1}\zeta_2^{m_2}\,$ and 
$\,\zeta^n\equiv\zeta_1^{n_1}\zeta_2^{n_2}$.
\,If the vectors $\,m\,$ and $\,n\,$ are parallel, then
$\,\varphi_{m,n}\,$ takes values in a circle in $\,G\,$ and, for
dimensional reasons, the identity (\ref{mntr}) still holds.
The integral homology group $\,H_2(G)\cong H_2(Z)\cong\NZ_2\,$ is 
generated by $\,\varphi_{(1,0),(0,1)}\,$ \cite{FGK}. \,It is easy to see 
from the local expression for the gerbe holonomy that
\qq
\CH_{\CG_k}(\varphi_{(1,0),(0,1)})^2\ =\ \CH_{\CG_k}(\varphi_{(2,0),(0,1)})\ 
=\ (-1)^k\,.
\label{sqr}
\qqq
The latter equality is obtained by noting that there exists a smooth 
map
$\,\tilde g:D\rightarrow G\,$ defined on the unit disc $\,D\,$ such
that $\,\,\tilde g(0)=1\,\,$ and
$\,\,\tilde g(\ee^{i\sigma_1})=\ee^{2i\sigma_1\theta_1}\,$ so that
$\,\,\tilde g(t\,\ee^{i\sigma_1\theta_1})\,\ee^{i\sigma_2\theta_2}\,\,$ 
provides the homotopy between $\,\varphi_{(0,0),(0,1)}\,$ and 
$\,\varphi_{(2,0),(0,1)}$. \,The use of this homotopy leads via 
Eq.\,(\ref{holcurv})  to the left equality in (\ref{sqr})
\cite{FGK}. We infer that
\qq
\CH_{\CG_k}(\varphi_{(1,0),(0,1)})\ =\ \pm\ee^{\pi i\,k/2}\,.
\nonumber
\qqq
Different choices of the sign correspond to the holonomy of gerbes
$\,\CG_k\,$ in two different isomorphism classes. \,The 
invariance (\ref{diffinv}) of the gerbe holonomy under the
orientation-preserving diffeomorphisms of $\,\NT^2\,$ implies
that 
\qq
\CH_{\CG_k}(\varphi_{m,n})\ =\ \CH_{\CG_k}(\varphi_{\tilde m,\tilde n})
\ \qquad{\rm if}\qquad (\tilde m,\tilde n)=(\begin{matrix}
{}_a&\hspace{-0.2cm}_b\cr{}^c&\hspace{-0.2cm}^d\end{matrix})(m,n)
\label{tildmn}
\qqq
for $\,(\begin{matrix}{}_a&\hspace{-0.2cm}_b\cr{}^c&\hspace{-0.2cm}^d
\end{matrix})\in SL(2,\NZ)\,$ 
(where $\,m,n,\tilde m,\tilde n\,$ are treated as column vectors). 
Using this invariance, it is easy to see that 
\qq
\CH_{\CG_k}(\varphi_{m,n})\ =\ (\pm\ee^{\pi i\,k/2})^{m\wedge n}
\nonumber
\qqq 
whenever the components of $\,m\,$
and $\,n\,$ are $\,0\,$ or $\,1\,$for $\,m\wedge n=m_1n_2-m_2n_1$. 
\,Employing similar homotopies as before, one may check that this 
equation remains true for all $\,m,\,n\in\NZ^2$.
\,Finally, the definition (\ref{cgg}) gives the result:
\qq
c_{(\zeta^{m},\zeta^{n}),
(\zeta^{m'},\zeta^{n'})}
\ =\ (\pm\ee^{\pi ik/2})^{m\wedge n'+m'\wedge n}\,\ee^{\pi i\,k\,
\tr(m'\theta\,n\theta\,-\,m\theta\,n'\theta)}\ \,\equiv\,\ c_{(m,n),(m',n')}.
\label{czzzz}
\qqq
It is straightforward to verify directly using Eqs.\,(\ref{cond'})
that the right hand side depends on the classes of $\,m,\,n,\,m',\,n'\,$
in $\,\NZ^2/2\NZ^2$. \,The $\,SL(2,\NZ)\,$ symmetry extends 
to the 2-cocycle $\,c_{(m,n),(m',n')}\,$ implying that
\qq
c_{(m,n),(m',n')}\ =\ c_{(\tilde m,\tilde n),(\tilde m',\tilde n')}
\nonumber
\qqq
if the pairs $\,(m,n)\,$ and $\,(\tilde m,\tilde n)\,$
are related as in (\ref{tildmn}). \,It is easy to check that 
$\,c_{(m,n),(m',n')}\equiv1\,$ if and only if the upper sign is chosen
on the right hand side of Eq.\,(\ref{czzzz}) and $\,k\in2\NZ\,$ if $\,r\,$
is even or $\,k\in4\NZ\,$ when $\,r\,$ is odd.
\,Note that the expression (\ref{czzzz}) for $\,c_{(m,n),(m',n')}\,$ 
encompasses also the formula (\ref{czz}) if we set $\,m\wedge n\equiv0\,$ 
for $\,Z=\NZ_N$. 

\smallskip

Summarizing, \,the obstruction FGK 2-cocycle 
$\,c_{(m,n),(m',n')}\,$ on $\,Z^2\,$ is given by Eq.\,(\ref{czzzz}).
We have included a table in Sec.\,\ref{sec:concl} listing those values 
of $k$ for which the FGK cocycle is trivial.

\section{FGK Cocycle and the Cohomological Obstruction}
\label{sec:cohFGK}

\no The obstruction cohomology class $\,\kappa\in H^2(G^2,U(1))
\cong \mathrm{Hom}(H_2(G^2),U(1))\,$ that corresponds to the  
isomorphism class of the flat gerbe $\,\CK_k\,$ over $\,G^2\,$
defined by (\ref{flatg}) may be easily described explicitly. Indeed, it 
assigns to fields $\,\varphi_1\times \varphi_2:\Sigma\rightarrow G^2\,$
inducing the homology classes $\,[\varphi_1\times \varphi_2]\in H_2(G^2)\,$ 
their holonomy with respect to the gerbe $\,\CK_k\,$:
\qq
\big\langle\,[\varphi_1\times\varphi_2]\,,\,\kappa\,\big\rangle\ 
=\ \CH_{\CK_k}(\varphi_1\times\varphi_2)\ =\ c_{\varphi_1,\varphi_2}\,,
\label{kappa}
\qqq
see Eq.\,(\ref{cgg}). \,In order to describe $\,\kappa$, \,it is then 
enough to calculate $\,c_{\varphi_1,\varphi_2}\,$ for fields 
$\,\varphi_{1,2}\,$ 
such that $\,[\varphi_1\times \varphi_2]\,$ generate $\,H_2(G^2)$.
\vskip 0.2cm

Let us first consider the case with $\,Z=\NZ_N$. \,Here 
\qq
H_2(G^2)\ \cong\ H_1(G)\otimes H_1(G)\ \cong\ \NZ_N\,,
\nonumber
\qqq
see (\ref{H222}) and (\ref{H22}), with the first isomorphism given by the 
cross product. The group $\,H_1(G)\cong Z\,$ is composed of the homology
classes of the maps
\qq
S^1\,\ni\,\ee^{i\sigma}\,\longmapsto\,\ee^{\,i\sigma m\theta}\,\in\,G
\nonumber
\qqq
that correspond to elements $\,\zeta^m\in Z\,$ for $\,\zeta
=\ee^{2\pi i\,\theta}$.
\,Consequently, $\,H_2(G^2)\,$ is composed of the
cross products of the latter classes. These are the homology classes 
of the maps 
\qq
\NT^2\,\ni\,(\ee^{i\sigma_1},\ee^{i\sigma_2})\,\longmapsto\,
(\ee^{\,i\sigma_1m\theta},\,\ee^{\,i\sigma_2n\theta})\,\in\,G^2\,,
\nonumber
\qqq
i.e. of $\,\varphi_{m,0}\times\varphi_{0,n}\,$ in the notation of
Eq.\,(\ref{27}). \,Eqs.\,(\ref{kappa}) and (\ref{czz}) give for the paring
of these classes with the cohomology class $\,\kappa\,$ the result:
\qq
\big\langle\,[\varphi_{m,0}\times\varphi_{0,n}]\,,\,\kappa\,\big\rangle\ 
=\ c_{\varphi_{m,0},\varphi_{0,n}}\ =\ \ee^{-\pi i\,k\,mn\,\tr\,\theta^2}. 
\nonumber
\qqq
The right hand side induces a bihomomorphism
$\,\xi_\kappa:Z^2\rightarrow U(1)$, 
\qq
\xi_\kappa(\zeta^m,\zeta^n)\ =\ \ee^{-\pi i\,k\,mn\,\tr\,\theta^2}.
\label{biho1}
\qqq
The discussion of Sec.\,\ref{sec:cohob} relating bihomomorphisms
to the cohomology classes, \,see Eq.\,(\ref{zZZ}), permits to identify 
$\,\kappa\,$ with the class in $\,H^2(Z^2,U(1))\cong 
H^2(G^2,U(1))\,$ generated by the 2-cocycle 
\qq
\chi_{(\zeta^m,\zeta^n),(\zeta^{m'},\zeta^{n'})}\ 
=\ \ee^{-\pi i\,k\,mn'\,\tr\,\theta^2}
\nonumber
\qqq
on the group $\,Z^2$. 

\vskip 0.2cm

Let us pass now to the case with $\,Z=\NZ_2^2$. \,Here
\qq
H_2(G^2)\ \cong\ H_2(G)\oplus H_1(G)\otimes H_1(G)\oplus H_2(G)
\ \cong\ \NZ_2^6\,,
\nonumber
\qqq
see the results (\ref{H222}) and (\ref{H22}). \,The first (resp. second) 
copy of $\,H_2(G)\cong\NZ_2\,$ injects into $\,H_2(G^2)\,$ to the homology 
classes of the fields
$\,\varphi_{m,n}\times \varphi_{0,0}\,$ for $\,m,\,n\in\NZ^2\,$ (resp. of the 
fields $\,\varphi_{0,0}\times \varphi_{m,n}$). \,The holonomy of gerbe 
$\,\CK_k\,$ is trivial along such fields so that
\qq
\big\langle\,[\varphi_{m,n}\times \varphi_{0,0}]\,,\,\kappa\,\big\rangle\ =\ 
c_{\varphi_{mn},\varphi_{0,0}}\ =\ 1\ =\ c_{\varphi_{0,0},\varphi_{m,n}}\ 
=\ \big\langle\,[\varphi_{0,0}\times \varphi_{m,n}]\,,\,\kappa\,\big\rangle\,.
\nonumber
\qqq
On the other hand, similarly as before, $\,H_1(G)\otimes H_1(G)\,$
injects to the homology classes in $\,H_2(G^2)\,$ of the fields
$\,\varphi_{m,0}\times\varphi_{0,n}\,$ so that
\qq
\big\langle\,[\varphi_{m,0}\times\varphi_{0,n}]\,,\,\kappa\,\big\rangle\ =\ 
c_{\varphi_{m,0},\varphi_{0,n}}\ =\ (\pm\ee^{\pi i\,k/2})^{m\wedge n}\,
\ee^{-\pi i\,k\,\tr\,m\theta\,n\theta},
\nonumber
\qqq 
see Eq.\,(\ref{czzzz}). Again, the right hand side induces a bihomomorphism
$\,\xi_\kappa:Z^2\rightarrow U(1)$,
\qq
\xi_\kappa(\zeta^m,\zeta^n)\ =\ (\pm\ee^{\pi i\,k/2})^{m\wedge n}\,
\ee^{-\pi i\,k\,\tr\,m\theta\,n\theta}\,.
\label{biho2}
\qqq
It permits to identify $\,\kappa\,$ with the cohomology class in 
$\,H^2(Z^2,U(1))\cong H^2(G^2,U(1))\,$
generated by the 2-cocycle 
\qq
\chi_{(\zeta^m,\zeta^n),(\zeta^{m'},\zeta^{n'})}\ =\ 
(\pm\ee^{\pi i\,k/2})^{m\wedge n'}
\,\ee^{-\pi i\,k\,\tr\,m\theta\,
n'\theta}
\nonumber
\qqq
on $\,Z^2$, \,see again Eq.\,(\ref{zZZ}). \,Note that the formula (\ref{biho2})
encompasses also the expression (\ref{biho1}) if, as before, we set 
$\,m\wedge n\equiv0\,$ for $\,Z=\NZ_N\,$ and that the bihomomorphisms 
$\,\xi_\kappa\,$ satisfy the relation
\qq
\xi_\kappa(\zeta^m,\zeta^m)\ =\ \ee^{-\pi i\,k\,\tr\,(m\theta)^2}.
\label{diag}
\qqq
\vskip 0.1cm

It is well known that the elements $\,z\in Z\,$ correspond
to simple currents $\,J_{z}\,$ of the level $\,k\,$ WZW theory \cite{SY},
\,i.e. to primary fields that induce under fusion with other primary
fields a permutation of the latter. The conformal weights $\,\Delta_z\,$ 
of the primary fields $\,J_z\,$ satisfy the relation
\qq
\Delta_{\zeta^m}\,=\,\frac{1}{2}k\,\tr\,(m\theta)^2\ \,{\rm mod}\ 1\,.
\nonumber
\qqq
The conditions (\ref{condN}) for $\,Z=\NZ_N\,$ or (\ref{cond'}) for
$\,Z=\NZ_2^2\,$ are equivalent to the requirement that the simple currents
$\,J_z\,$ for $\,z\in Z\,$ be effective (in the terminology of \cite{GRS}),
i.e. that
\qq
N_z\Delta_z\,\in\,\NZ\qquad{\rm for}\qquad z\in Z
%\label{cond''}
\nonumber
\qqq
where $\,N_z\,$ stands for the order of the element $\,z$. 
\,Eq.\,(\ref{diag}) becomes the identity
\qq
\xi_\kappa(z,z)\ =\ \ee^{-2\pi i\,\Delta_z}.
%\label{KSB}
\nonumber
\qqq 
Bihomomorphisms with the above property on arbitrary groups 
of effective simple currents have been studied in the context of 
simple-current orbifolds of conformal field theories in \cite{KS}.
\,In \cite{FRS} they were called the Kreuzer-Schellekens (KS) 
bihomomorphisms. Note that if 
\qq
\Delta_z\,\in\,\NZ\qquad{\rm for}\qquad z\in Z
\label{cond'''}
\qqq
then $\,\xi_\kappa(z,z)=1$. \,Such bihomomorphisms are called alternating.
They are in one-to-one correspondence, see Lemma 3.16 of \cite{FRS}, 
with the cohomology classes in $\,H^2(Z,U(1))$. \,The latter group is 
trivial for $\,Z=\NZ_N\,$
and in this case the condition (\ref{cond'''}) assures the triviality 
of the KS bihomomorphism $\,\xi_\kappa$. \,For $\,Z=\NZ_2^2$, \,however,
$\,H^2(Z,U(1))=\NZ_2\,$ and even if the condition (\ref{cond'''})
is satisfied, the KS bihomomorphism $\,\xi_\kappa\,$ may be non-trivial
which indeed happens for the choice of the gerbe $\,\CG_k\,$ corresponding
to the lower sign on the right hand side of Eq.\,(\ref{biho2}). 
\vskip 0.1cm

As we have shown, there is a close relation between the cohomology
class $\,\kappa\,$ obstructing the existence of a stable
isomorphism (\ref{2grbs}) and the FGK obstruction 2-cocycle 
$\,c\,$ on $\,Z^2\,$ obtained
from the generalized Polyakov-Wiegmann formula. The cohomology class 
$\,\kappa\,$ comes from the KS bihomomorphism
$\,\xi_\kappa:Z^2\rightarrow U(1)\,$ of Eq.\,(\ref{biho2}) via the embedding
\qq
\mathrm{Hom}(Z\otimes Z,U(1))\,\hookrightarrow\,H^2(Z^2,U(1))
\ \cong\ H^2(G^2,U(1))
\label{emb}
\qqq
with the first arrow mapping $\,\xi_\kappa\,$ into the cohomology class
of the 2-cocycle
\qq
\chi_{(z_1,z_2),(z'_1,z'_2)}\ =\ \xi_\kappa(z_1,z'_2)\,.
\nonumber
\qqq
On the other hand, the FGK 2-cocycle on $\,Z^2\,$ has the form
\qq
c_{(z_1,z_2),(z'_1,z'_2)}\ =\ \xi_\kappa(z_1,z'_2)
\ \xi_\kappa(z_2,z'_1)^{-1}\,.
\nonumber
\qqq
The triviality of the obstruction cohomology class 
$\,\kappa\,$ generated by $\,\chi_\kappa\,$ must be 
equivalent to the triviality of the KS bihomomorphism 
$\,\xi_\kappa$. \,This may be also checked by a direct calculation.
On the other hand, the triviality of the bihomomorphism $\,\xi_\kappa\,$
is clearly equivalent to that of the FGK 2-cocycle $\,c$. 
This establishes the equivalence of three different presentations of
the obstruction. 
Note, for example, that in the case with $\,Z=\NZ_2$, \,the bihomomorphism 
$\,\xi_\kappa\,$ given by Eq.\,(\ref{biho1}) is non-trivial if 
$\,k\,\tr\,\theta^2\,$ is an odd integer and the corresponding 2-cocycle 
$\,\chi\,$ is  cohomologically non-trivial whereas, as discussed above, 
the FGK 2-cocycle $\,c\,$ is non-trivial but cohomologically trivial.

\section{Equivariant Multiplicative  Gerbes} 
\label{sec:emgerbes}

\noindent In this section we shall define multiplicative and 
equivariant-multiplicative structures on gerbes over Lie groups $G$. 
Some preliminary notations will be needed.
First we recall that the sequence $\left \lbrace G^p \right \rbrace$ 
of powers of $G$ forms a  simplicial manifold. Here we only need one 
aspect of this assertion, namely that there are ``face maps'' 
$\Delta_k^p: G^{p} \to G^{p-1}$ for all $p> 1$ and $0\leq k\leq p$, namely
\qq
%\label{facemaps}
\nonumber
\Delta_k^{p}(g_1,...,g_p) := \left\lbrace\begin{array}{ll}
(g_2,...,g_p) & \text{ for } k=0 \\
(g_1,...,g_{k-1}g_k,...,g_p) & \text{ for } 1 \leq k < p \\
(g_1,...,g_{p-1}) & \text{ for }k=p\text{,} \\
\end{array} \right. 
\qqq
and that these face maps satisfy the simplicial relations
\qq
\label{simprel}
\Delta^{p-1}_h \circ \Delta^p_k = \Delta^{p-1}_{k-1} \circ \Delta^{p}_h
\qqq
for all $h<k$. 
Such a structure is also called an ``incomplete'' simplicial manifold. 
Notice that the group multiplication $m:G \times G \to G$ and the 
projections $p_{1,2}:G \times G \to G$ can be rediscovered as 
$\Delta^2_1 = m$, $\Delta^2_2=p_1$ and $\Delta^2_0 = p_2$. We will 
sometimes suppress the upper index of $\Delta^p_k$.
A differential form $\omega\in\Lambda^n(G^2)$ 
will be called multiplicative, if 
\qq
\label{rhodelta}
\Delta_0^{*}\omega + \Delta_2^{*}\omega = \Delta_3^{*}\omega + 
\Delta_1^{*}\omega\text{.}
\qqq
In this case we denote the  $n$-form \erf{rhodelta} by $\omega_{\Delta}$. 
\smallskip

Multiplicative structures are considered for pairs $(\mathcal{G},\omega)$ 
composed of a gerbe $\mathcal{G}$ over $G$ with curvature $\,H\,$
and a multiplicative 2-form $\omega \in \Lambda^2(G^2)$ satisfying
\qq
m^{*}H = p_1^{*}H + p_2^{*}H + \mathrm{d}\omega\text{.}
\label{3Hnew}
\qqq
Our main example will involve the pair $\,(\CG_k,\omega_k)\,$ composed of 
a gerbe with the curvature 3-form $\,H_k\,$ of Eq.\,(\ref{meg}) and of 
the 2-form $\,\omega_k\,$ of Eq.\,(\ref{rho}). In this case, 
the identity \erf{3Hnew} 
is shown by Eq.\,\erf{3H} and $\omega_k$ is multiplicative due to Eq. 
\erf{omegadeltaclosed}. \,A multiplicative structure \cite{WMult} on 
$(\mathcal{G},\omega)$ is a 1-isomorphism
\qq
\mathcal{M} : m^{*}\mathcal{G} \to p_1^{*}\mathcal{G} \otimes p_2^{*}
\mathcal{G} \otimes \mathcal{I}_{\omega}
\label{multmorph}
\qqq
of gerbes over $G^2$ and a 2-isomorphism 
\qq
\label{alpha}
\alpha: (\id \otimes \Delta_0^{*}\mathcal{M} \otimes \id) \circ 
\Delta_2^{*}\mathcal{M} \Rightarrow (\Delta_3^{*}\mathcal{M} \otimes 
\id \otimes \id) \circ \Delta_1\mathcal{M} \qqq
between 1-isomorphisms of gerbes
over $G^3$ which satisfies a natural pentagon axiom over $G^4$. The condition 
that $\omega$ is multiplicative is required for the existence of $\alpha$.
In the particular case of the pair $\,(\CG_k,\omega_k)$, \,the isomorphism 
$\mathcal{M}$ is the isomorphism (\ref{2grbs}) in Sec.\,\ref{sec:cohob}. 
Two multiplicative gerbes $(\mathcal{G}^a,\mathcal{M}^a,\alpha^a)$ and 
$(\mathcal{G}^b,\mathcal{M}^b,\alpha^b)$ are equivalent, if there exists an 
isomorphism $\mathcal{B}:\mathcal{G}^a \to \mathcal{G}^b$ and a 2-isomorphism 
\qq
\beta: (p_1^{*}\mathcal{B} \otimes p_2^{*}\mathcal{B} \otimes \id) \circ 
\mathcal{M}^a \Rightarrow \mathcal{M}^b \circ m^{*}
\mathcal{B}
\label{equivmult}
\qqq
which is compatible with $\alpha^a$ and $\alpha^b$ in a certain way 
\cite{WMult}. 
Equivalent multiplicative gerbes have the same
curvature 3-form $\,H\,$ and the same multiplicative 2-form $\,\omega$.
\smallskip

Next we combine a multiplicative structure on a gerbe $\mathcal{G}$ 
with an equivariant structure. 
In general, if a discrete group $Z$ acts smoothly on the left on a manifold 
$M$ over which a gerbe $\mathcal{G}$ is defined, a $Z$-equivariant structure 
on $\mathcal{G}$ \cite{GSW} consists of a collection of isomorphisms
\qq
\mathcal{A}_{z}: \mathcal{G}
\to z\mathcal{G}\text{,}
\nonumber
\qqq
where $z\mathcal{G} := (z^{-1})^{*}\mathcal{G}$, and a collection 
of 2-isomorphisms
\qq
\varphi_{z_1,z_2}: z_1\mathcal{A}_{z_2}
\circ \mathcal{A}_{z_1}
\Rightarrow \mathcal{A}_{z_1z_2}
\nonumber
\qqq
such that the diagram
\qq
\label{5}
\alxydim{@C=2cm@R=1cm}{z_1 z_2\mathcal{A}_{z_3} \circ
z_1\mathcal{A}_{z_2}
\circ \mathcal{A}_{z_1} \ar@{=>}[d]_{z_1\varphi_{z_2,z_3}
 \circ \id} \ar@{=>}[r]^-{\id \circ \varphi_{z_1,z_2}}
& z_1 z_2\mathcal{A}_{z_3}
\circ \mathcal{A}_{z_1z_2} \ar@{=>}[d]^{\varphi_{z_1z_2,z_3}}
\\ z_1\mathcal{A}_{z_2z_3} \circ
\mathcal{A}_{z_1}
\ar@{=>}[r]_-{\varphi_{z_1,z_2z_3}}
& \mathcal{A}_{z_1z_2z_3}}
\qqq
of 2-isomorphisms is commutative. 
We need the following facts:
\begin{enumerate}
\item 
Suppose that we have two manifolds $M_1$ and $M_2$ with smooth left 
actions of discrete groups $Z_1$ and $Z_2$, respectively.  Suppose further 
that  $\varphi: Z \to Z'$ is a group homomorphism and that $f:M_1 \to M_2$ 
is a smooth map that exchanges the actions in the sense that
\qq
\label{mapcomp}
f(z x) = \varphi(z)f(x) 
\qqq
for all $x\in M_1$ and $z\in Z_1$. Then, the pullback $f^{*}\mathcal{G}$ of a 
$Z_2$-equivariant bundle gerbe over $M_2$ carries a canonical 
$Z_1$-equivariant structure. 

\item
In the case when $Z$ acts freely and properly
on $M$, the quotient $M/Z$ is a smooth manifold and the projection 
$p: M \to M/Z$ is a surjective submersion. 
The pullback $p^{*}\mathcal{G}$ of a  gerbe $\mathcal{G}$ over $M/Z$ carries 
a canonical $Z$-equivariant structure. Conversely, every gerbe $\mathcal{G}$ 
over $M$ with a $Z$-equivariant structure defines a ``descent'' gerbe 
$\mathrm{Des}_Z(\mathcal{G})$ over $M/Z$. These two procedures are inverse to 
each other in an appropriate sense, see \cite{GSW}.  

\item
Suppose that we have two smooth manifolds $M_1$ and $M_2$, both with free 
and proper 
left actions of discrete groups $Z_1$ and $Z_2$, respectively. Given a group 
homomorphism 
$\varphi: Z_1 \to Z_2$ and a smooth map $f: M_1 \to M_2$ satisfying 
\erf{mapcomp}, there exists a unique map $g: M_1/Z_1 \to M_2/Z_2$ between the 
quotients such that $p_2 \circ f= g \circ p_1$. Then, 
\qq
\label{despull}
\mathrm{Des}_{Z_1}\circ f^{*} = g^{*} \circ \mathrm{Des}_{Z_2}\text{.}
\qqq
Thus, descent is compatible with pullbacks. It is also compatible with 
tensor products.
\end{enumerate}
We further need the definition of $Z$-equivariant isomorphisms and 
2-isomorphisms. For an isomorphism being $Z$-equivariant is not a property 
but additional structure. A $Z$-equivariant structure on an isomorphism 
$\mathcal{B}:\mathcal{G}^a \to \mathcal{G}^b$ between gerbes with 
$Z$-equivariant structures $(\mathcal{A}^a_z,\varphi^a_{z_1,z_2})$ and 
$(\mathcal{A}^b_z,\varphi^b_{z_1,z_2})$ is a 2-isomorphism 
\qq
\eta_{z}: z\mathcal{B} \circ \mathcal{A}^a_{z} \Rightarrow 
\mathcal{A}^b_{z} \circ \mathcal{B}
\nonumber
\qqq
such that the diagram
\qq
\label{17}
\alxydim{@C=3cm@R=1cm}{z_1 z_2\mathcal{B} \circ z_1\mathcal{A}^a_{z_2} 
\circ \mathcal{A}^a_{z_1} \ar@{=>}[d]_-{z_1\eta_{z_2} \circ 
\id_{\mathcal{A}^a_{z_1}}} \ar@{=>}[r]^-{\id_{z_1 z_2\mathcal{B}} 
\circ \varphi^a_{z_1,z_2}} & z_1z_2\mathcal{B} \circ \mathcal{A}^a_{z_1z_2} 
\ar@{=>}[dd]^-{\eta_{z_1z_2}} \\ z_1\mathcal{A}_{z_2}^{b}\circ z_1 
\mathcal{B}\circ \mathcal{A}^a_{z_1} \ar@{=>}[d]_-{\id_{z_1
\mathcal{A}_{z_2}^b}\circ \eta_{z_1}} & \\ z_1\mathcal{A}^{b}_{z_2}
\circ \mathcal{A}^{b}_{z_1} \circ \mathcal{B} \ar@{=>}[r]_
-{\varphi^b_{z_1,z_2} \circ \id_{\mathcal{B}}} & \mathcal{A}^b_{z_1z_2} 
\circ \mathcal{B}}
\qqq
of 2-isomorphisms is commutative. Finally, a $Z$-equivariant 2-isomorphism 
$\phi: \mathcal{B}\Rightarrow \mathcal{B}'$ between isomorphisms 
$\mathcal{B}$ and $\mathcal{B}'$ with $Z$-equivariant structures $\eta_z$ 
and $\eta_z'$, respectively, is called $Z$-equivariant, if the diagram
\qq
\label{35}
\alxydim{@=1cm}{z\mathcal{B} \circ
\mathcal{A}^a_{z} \ar@{=>}[d]_{z\phi \circ
\id_{\mathcal{A}^a_{z}}} \ar@{=>}[r]^-{\eta_{z}} &
\mathcal{A}^b_{z} \circ \mathcal{B}
\ar@{=>}[d]^{\id_{\mathcal{A}^b_{z}} \circ \phi} \\
z\mathcal{B}' \circ \mathcal{A}^a_{z}
\ar@{=>}[r]_{\eta_{z}'} & \mathcal{A}^b_{z} \circ
\mathcal{B}'}
\qqq
of 2-isomorphisms is commutative. In case of a free and proper
group action, equivariant isomorphisms and 2-isomorphisms descent to 
the quotient in a way compatible with pullbacks and tensor products. 

\smallskip

In order to combine a multiplicative structure with an equivariant structure, 
the action $\rho: Z \times G \to G$ has to be compatible with the group 
multiplication of $G$ in the sense that $\rho$ is a group homomorphism. 
We will call such group actions ``multiplicative''. Multiplicative groups 
actions have the following two properties: if we let $Z^p$ act component-wise 
on $G^p$, the face maps  $\Delta^p_k: G^{p+1} \to G^p$ introduced above 
satisfy condition \erf{mapcomp}, where the group homomorphism $\varphi: 
Z^p \to Z^{p-1}$ is given by the  face map $\Delta_k^p$ of the group $Z$. 
In other words, there are commutative diagrams
\qq
\label{multact}
\alxydim{@C=1.5cm}{G^p \ar[r]^{z}\ar[d]_{\Delta_k^p} & G^p 
\ar[d]^{\Delta^p_k} \\ G^{p-1} \ar[r]_-{\Delta_k^p(z)} & G^{p-1}}
\qqq
for all $p$ and $0 \leq k\leq p$.  This property of the action $\rho$ 
guarantees, for instance, that if $\mathcal{G}$ is a $Z$-equivariant gerbe 
over $G$, the pullbacks $m^{*}\mathcal{G}$, $p_1^{*}\mathcal{G}$ and 
$p_2^{*}\mathcal{G}$ are $Z^2$-equivariant gerbes over $G^2$. The second 
property of a multiplicative group action is that in case of a free and 
proper group action, in which the quotient $G/Z$ is again a Lie group, 
the projection $p: G \to G/Z$ is a Lie group homomorphism. Most importantly, 
all of this holds for $Z$ a subgroup of the center of $G$ acting by 
multiplication.

\smallskip

Given a multiplicative group action, equivariant multiplicative structures 
are considered for pairs $(\mathcal{G},\omega)$ of a gerbe $\mathcal{G}$ 
over $G$ and a multiplicative, $Z^2$-invariant 2-form $\omega\in\Lambda^2(G^2)$
satisfying \erf{3Hnew} as before. Notice that such 2-forms define 
$Z^2$-equivariant trivial bundle gerbes $\mathcal{I}_{\omega}$. 
We say that a $Z$-multiplicative structure on $(\mathcal{G},\omega)$ is a 
$Z$-equivariant structure $(\mathcal{A}_z,\varphi_{z_1,z_2})$ on 
$\mathcal{G}$, a $Z^2$-equivariant isomorphism $(\mathcal{M},\eta_{z_1,z_2})$ 
like in \erf{multmorph} and a $Z^3$-equivariant 2-isomorphism $\alpha$ like 
in \erf{alpha}, satisfying the pentagon axiom. Two $Z$-multiplicative gerbes 
are equivalent, if there exists a $Z$-equivariant isomorphism 
$(\mathcal{B},\kappa_{z}):\mathcal{G}^a \to  \mathcal{G}^b$ and a 
$Z^2$-equivariant 2-isomorphism $\beta$ like in \erf{equivmult}, satisfying 
the same compatibility condition. 

\smallskip

The purpose of  $Z$-multiplicative gerbes over $G$ is that they correspond, 
for a free and proper 
group action, to multiplicative gerbes over the quotient $G' = G/Z$. This 
follows from the properties of equivariant structures listed above: the 
$Z$-equivariant  gerbe $\mathcal{G}$ determines a bundle gerbe 
$\mathcal{G}':= \mathrm{Des}(\mathcal{G})$ over $G'$. Eq.\,\erf{despull} 
implies 
that $m^{*}_{12}\mathcal{G}' = \mathrm{Des}(m^{*}\mathcal{G})$, and similarly, 
$p_i^{*}\mathcal{G}' = \mathrm{Des}(p_i^{*}\mathcal{G})$ for $i=1,2$. Further, 
the $Z^2$-equivariant, multiplicative 2-form $\omega$ determines a 2-form 
$\omega' \in \Lambda^2(G'^2)$, and this 2-form is again multiplicative.
Thus, the $Z^2$-equivariant 1-isomorphism $\mathcal{M}$ determines a 
1-isomorphism
\qq
\mathcal{M}' := \mathrm{Des}(\mathcal{M}): m^{*}\mathcal{G}' \to 
p_1^{*}\mathcal{G}' \otimes p_2^{*}\mathcal{G}' \otimes 
\mathcal{I}_{\omega'}\text{.}
\nonumber
\qqq
In the same way, the $Z^3$-equivariant 2-isomorphism $\alpha$ determines 
a 2-isomorphism $\alpha'$ as required for a multiplicative  gerbe over $G'$. 
This 2-isomorphism $\alpha'$ automatically satisfies the pentagon axiom. 
Thus every $Z$-multiplicative  gerbe over $G$ determines a multiplicative  
gerbe over the the quotient $G'$. In the same way,  equivalent 
$Z$-multiplicative gerbes determine equivalent multiplicative gerbes 
over $G'$.\vskip 0.1cm 

Summarizing, if $Z$ is a discrete group acting on the left on a Lie group 
$G$ in a smooth, multiplicative and free and proper way, we have a bijection  
between 
equivalence classes of $Z$-multiplicative gerbes over $G$ and equivalence 
classes of multiplicative gerbes over $G/Z$. The goal of the following 
sections is to classify $Z$-multiplicative structures on the pairs 
$(\mathcal{G}_k,\omega_k)$  over all compact, simple and simply-connected 
Lie groups $G$, for $Z$ a subgroup of the center of $G$.

\section{Local  Description of Equivariant Multiplicative Gerbes}
\label{sec:local}

\no
In this section we connect the geometrical definition of equivariant 
multiplicative gerbes to the cohomological language used in the first five 
sections. The cohomology theory that is most appropriate for gerbes, 
i.e. hermitian bundle gerbes with unitary connection, is the (real) 
Deligne cohomology. We shall recall some basic facts about it 
\cite{top,Gaj,GSW}.
\smallskip

Let us first consider a general manifold $M$. We denote  by $\mathcal{U}$ the
sheaf of smooth $U(1)$-valued functions, and by $\Lambda^{q}$
the sheaf of $q$-forms. For $\mathfrak{O}$ an open cover of $M$, 
the Deligne cohomology $\mathbb{H}^n(\mathfrak{O},\mathcal{D}(2))$ is 
the cohomology of the complex
\qq
%\label{150a}
\nonumber
\alxydim{@C=1cm}{0 \ar[r] & A^0(\mathfrak{O}) \ar[r]^-{D_0}
& A^1(\mathfrak{O}) \ar[r]^{D_1} & A^2(\mathfrak{O}) 
\ar[r]^{D_2} & A^3(\mathfrak{O}) \ar[r]& ...}
\qqq
with the cochain
groups
\begin{eqnarray*}
A^0(\mathfrak{O}) &=& C^0(\mathfrak{O},\mathcal{U})\textrm{,} \\
A^1(\mathfrak{O}) &=& C^0(\mathfrak{O},\Lambda^1) 
\oplus C^1(\mathfrak{O},\mathcal{U})\textrm{,} \\
A^2(\mathfrak{O}) &=& C^0(\mathfrak{O},\Lambda^2) 
\oplus C^1(\mathfrak{O},\Lambda^1) \oplus C^2(\mathfrak{O},
\mathcal{U})\textrm{,} \\
A^3(\mathfrak{O}) &=& C^1(\mathfrak{O},\Lambda^2) \oplus 
C^2(\mathfrak{O},\Lambda^1) \oplus
C^3(\mathfrak{O},\mathcal{U})\textrm{.}
\end{eqnarray*}
Here, $C^\ell(\mathfrak{O},\mathcal{S})$ denotes the $\ell^{\rm th}$ \v Cech 
cochain group of the open cover $\mathfrak{O}$ with values in a sheaf 
$\mathcal{S}$ of Abelian groups.
The differentials are
\begin{eqnarray*}
D_0(f_i) &=& (-\mathrm{i}f_i^{-1}df_i\;,
\; f_j^{-1}\cdot f_i)\textrm{,} \\
D_1(\Pi_i,\chi_{ij}) &=& (d\Pi_i\;,\;
-\mathrm{i}\chi_{ij}^{-1}d\chi_{ij}+\Pi_{j}
- \Pi_{i}\;,\;\chi_{jk}^{-1}\cdot \chi_{ik}\cdot \chi_{ij}^{-1})\textrm{,} \\
D_2(B_i,A_{ij},g_{ijk}) &=& (dA_{ij}-B_{j} + B_i \;,\;
-\mathrm{i}g_{ijk}^{-1}dg_{ijk}+A_{jk}-A_{ik}+A_{ij}\;,\;
 g_{jkl}^{-1} \cdot g_{ikl} \cdot g_{ijl}^{-1}\cdot g_{ijk})\textrm{.}
\end{eqnarray*}
A refinement $r: \mathfrak{O}' \to \mathfrak{O}$ of open covers induces the 
restriction maps $\mathbb{H}^n(\mathfrak{O},\mathcal{D}(2)) \to 
\mathbb{H}^n(\mathfrak{O}',\mathcal{D}(2))$ turning the Deligne cohomology 
groups into a direct system of groups. Its direct limit is denoted 
$\mathbb{H}^n(M,\mathcal{D}(2))$. 

\smallskip

Let us briefly recall what local data of gerbes, isomorphisms and 
2-isomorphisms are, for the details we refer the reader to \cite{GSW}. 
For a given gerbe $\mathcal{G}$ over $M$, one can choose a sufficiently
``good'' open cover $\mathfrak{O}$ of $M$ that permits 
to extract a cocycle $c \in A^2(\mathfrak{O})$, $D_2c=0$, in a certain way. 
Suppose that two gerbes $\mathcal{G}_1$ and $\mathcal{G}_2$ are given, and 
$\mathfrak{O}_1$ and $\mathfrak{O}_2$ are open covers that permit
to extract cocycles
$c_1$ and $c_2$. Suppose further that $\mathcal{A}: \mathcal{G}_1 
\to \mathcal{G}_2$ is an isomorphism. Then one can choose a common 
refinement $\mathfrak{O}$ of $\mathfrak{O}_1$ and $\mathfrak{O}_2$ that 
permits to extract a cochain 
$b \in A^1(\mathfrak{O})$ such that $c_2 = c_1 + D_1b$. The cochains for 
isomorphisms add under the composition of these isomorphisms. Finally, 
if a 2-isomorphism $\varphi:\mathcal{A}_1 \Rightarrow \mathcal{A}_2$ is 
given and $b_1$ and $b_2$ are cochains for $\mathcal{A}_1$ and $\mathcal{A}_2$,
respectively, for a suitable open cover $\mathfrak{O}$, one can always 
extract a cochain $a\in A^0(\mathfrak{O})$ such 
that $b_2=b_1+D_0a$. The cochains for 2-isomorphisms add under both the 
horizontal and the vertical composition of 2-isomorphisms.
Conversely, one can reconstruct gerbes, isomorphisms and 2-isomorphisms 
from given local data, and the two procedures are inverse to each other 
in an appropriate sense. In particular, they establish a bijection between 
$\mathbb{H}^2(M,\mathcal{D}(2))$ and the set of isomorphism classes 
of gerbes over $M$.

\smallskip

In the following we want to apply the procedure of extraction of local data to 
an equivariant multiplicative gerbe. This requires a careful discussion of 
open covers $\mathfrak{O}^p=\left \lbrace O^p_i \right \rbrace{}_{i\in I^p}$ 
of powers $G^p$ of a Lie group $G$. As the definition of a $Z$-multiplicative 
gerbe over $G$ involves pullbacks along the face maps 
$\Delta^p_k: G^{p} \to G^{p-1}$ and along the action $\rho: Z \times G \to G$, 
we need the open covers to be compatible with all these maps. 

\smallskip

We say that a sequence $\left \lbrace \mathfrak{O}^p \right \rbrace$ of open 
covers of $G^p$ is ``simplicial'', if the sequence 
$\left \lbrace \mathcal{I}^p \right \rbrace$ of index sets forms an 
incomplete simplicial set (i.e. there are face maps  $\Delta_k^p: I^{p} 
\to I^{p-1}$ satisfying \erf{simprel}), such that
\qq
\label{simpcond}
\Delta_k^p(O^{p}_i) \subset O^{p-1}_{\Delta^p_k(i)}
\qqq
for all $p > 1$, all $0 \leq k \leq p$ and all $i\in I^p$. For a simplicial 
sequence of open covers one has induced chain maps
\qq
(\Delta_k^p)^{*}: C^{\ell}(\mathfrak{O}^{p-1},\mathcal{S}) 
\to C^{\ell}(\mathfrak{O}^p,\mathcal{S})
\quad\text{ defined by }\quad \left ( (\Delta_l^p)^{*}f \right )_i 
:= (\Delta_k^p)^{*}(f_{\Delta_k^p(i)})\text{,}
\label{liftfacemaps}
\qqq
satisfying the co-simplicial relations
\qq
(\Delta_k^p)^{*} \circ (\Delta_h^{p-1})^{*} = (\Delta_h^p)^{*} \circ 
(\Delta_{k-1}^{p})^{*}
\label{cosimplicial}
\qqq
for $h<k$. We further say that a sequence $\left \lbrace \mathfrak{O}^p 
\right \rbrace$ of open covers is ``$Z$-equivariant'', if each cover 
$\mathfrak{O}^p$ is $Z^p$-equivariant in the sense that its index set 
$I^p$ carries an action of $Z^p$ in such a way that
\qq
\label{equivcond}
z(O^p_i) \subset O^p_{zi}
\qqq
for all $z\in Z^p$ and $i\in I^p$. For a $Z$-equivariant sequence of open 
covers one has an induced action of $Z^p$ on $C^{\ell}(\mathfrak{O}^p,
\mathcal{S})$ by chain maps, namely
\qq
\label{liftgroupaction}
z: C^{\ell}(\mathfrak{O}^p,\mathcal{S}) \to C^{\ell}(\mathfrak{O}^p,
\mathcal{S})\quad\text{ defined by }\quad
(zf)_i := (z^{-1})^{*}(f_{z^{-1}(i)})
\qqq
for $z\in Z^p$ and $f \in C^{\ell}(\mathfrak{O}^p,\mathcal{S})$. Combining 
both notions, we say that a sequence $\left \lbrace \mathfrak{O}^p \right 
\rbrace$ of open covers is ``$Z$-simplicial'', if it is both simplicial and 
$Z$-equivariant, and if the face maps $\Delta_k^p:I^{p} \to I^{p-1}$ and 
the action of $Z^p$ on $I^{p}$ commute in the sense that all diagrams
\qq
\label{multact1}
\alxydim{@C=1.5cm}{I^p \ar[r]^{z}\ar[d]_{\Delta_k^p} & I^p 
\ar[d]^{\Delta^p_k} \\ I^{p-1} \ar[r]_-{\Delta_k^p(z)} & I^{p-1}}
\qqq
are commutative (cf. diagram \erf{multact}). This compatibility condition 
ensures that the induced maps \erf{liftfacemaps} and \erf{liftgroupaction} 
on the \v Cech cohomology groups commute in the same way, i.e.   
\qq
z \circ (\Delta_k^p)^{*} = (\Delta_k^p)^{*} \circ \Delta_k^p(z)\text{.}
\label{diffcommute}
\qqq

\smallskip

In the following we shall use $Z$-simplicial sequences of open covers 
to extract local data of $Z$-multiplicative gerbes. We have included 
an appendix (after Sec.\,\ref{sec:concl}) in which we prove the following. 
Suppose that a finite Abelian group $Z$ acts on $G$ in a smooth, 
multiplicative, free and proper way, and assume that 
$\left \lbrace \mathfrak{V}^p \right \rbrace$ is any sequence of open 
covers $\mathfrak{V}^p$ of $G^p$. Then, there exists a $Z$-simplicial 
sequence $\left \lbrace \mathfrak{O}^p \right \rbrace$ of open covers, 
such that each $\mathfrak{O}^p$ is a refinement of $\mathfrak{V}^p$. 
As a consequence, one can choose  open covers $\mathfrak{V}^p$ separately 
for all $p$, in such a way that they permit to extract local data 
of any given combination of gerbes and isomorphisms. Since each open 
cover $\mathfrak{O}^p$ of a $Z$-simplicial refinement is finer than 
$\mathfrak{V}^p$, also the new covers $\mathfrak{O^p}$ permit to extract 
local data of the given structure. We can hence assume that one can 
always choose sufficiently fine $Z$-simplicial sequences of open covers.

\smallskip

For a given $Z$-simplicial sequence $\left \lbrace \mathfrak{O}^p \right 
\rbrace$ of open covers, we  consider the groups 
\qq
K^{p,q,n} := \mathrm{Map}((Z^{p})^q,A^n(\mathfrak{O}^p))
\nonumber
\qqq
with elements denoted like $x_{z_1,...,z_q} \in A^n(\mathfrak{O}^p)$, for 
$z_1,...,z_q$ elements in $Z^p$. 
On the groups $K^{p,q,n}$ we find three operators: the first 
is the Deligne differential 
\qq
D_{p,q,n}: K^{p,q,n} \to K^{p,q,n+1}
\quad\text{ with }\quad
(D_{p,q,n}(x))_{z_1,...,z_q} := D_n(x_{z_1,...,z_q})\text{.}
\nonumber
\qqq
The second is the ``group cohomology differential of the 
group $Z^{p}$\hspace{0.025cm}''
\qq
\delta_{p,q,n}: K^{p,q,n} \to K^{p,q+1,n}
\quad\text{ with }\quad
(\delta_{p,q,n}(x))_{z_0,...,z_q} := z_0x_{z_1,...,z_q} - x_{z_0z_1,...,z_q} 
+ ... \pm x_{z_0,...,z_{q-1}}\text{,}
\nonumber
\qqq
whose definition uses the lift \erf{liftgroupaction} of the $Z^p$-action 
to the Deligne cochain group $A^n(\mathfrak{O}^p)$.
The third operator we have is the simplicial operator
\qq
\Delta_{p,q,n}: K^{p,q,n} \to K^{p+1,q,n}
\quad\text{ with }\quad
(\Delta_{p,q,n}(x))_{z_1,...,z_q} := \sum_{k=0}^{p+1} (-1)^{k} 
(\Delta_k^{p+1})^{*}(x_{\Delta^p_kz_1,...,\Delta^p_kz_q}) \text{,}
\label{simpop}
\qqq
whose definition uses the lift \erf{liftfacemaps} of the face maps to 
the Deligne cochain groups. Notice that in \erf{simpop}  $z_1,...,z_q$ are  
elements of $Z^{p+1}$ and $\Delta^p_k:Z^{p+1} \to Z^{p}$ is the face map 
of the group $Z$.  Due to the co-simplicial relations \erf{cosimplicial}, 
we have $\Delta_{p+1,q,n} \circ \Delta_{p,q,n}=0$. The Deligne differential 
$D$ commutes with  pullbacks, and thus with both operators $\delta$ and 
$\Delta$. Further, the differentials $\delta$ and $\Delta$ commute due to 
\erf{diffcommute}. This endows $K^{p,q,n}$ with the structure of a triple 
complex.

\smallskip

Now we are prepared to list local data of a $Z$-equivariant multiplicative 
gerbe over $G$. We chose a $Z$-simplicial sequence $\left \lbrace 
\mathfrak{O}^p \right \rbrace$ of open covers that permit to extract 
local data of $\mathcal{G}$ and all involved isomorphisms and 2-isomorphisms. 
Then, the $Z$-equivariant gerbe 
$(\mathcal{G},\mathcal{A}_{z},\varphi_{z_1,z_2})$ has local data
$c\in K^{1,0,2}$, $b\in K^{1,1,1}$ and $a \in K^{1,2,0}$
satisfying the relations
\qq
D_{1,0,2}c=0
\quad\text{, }\quad
\delta_{1,0,2} c = D_{1,1,1}b
\quad\text{, }\quad
\delta_{1,1,1} b = -D_{1,2,0} a
\quad\text{ and }\quad
\delta_{1,2,0} a=0\,\text{,}
\nonumber
\qqq
of which the last one is the commutativity of diagram \erf{5}, see \cite{GSW}.
The $Z^2$-equivariant isomorphism $(\mathcal{M},\eta_{z_1,z_2})$ has local data
$\beta\in K^{2,0,1}$ and $\phi\in K^{2,1,0}$
satisfying
\qq
\Delta_{1,0,2}c + \omega = D_{2,0,1}\beta
\,\text{, }\quad
\delta_{2,0,1} \beta  = \Delta_{1,1,1} b- D_{2,1,0} \phi
\quad\text{ and }\quad
\Delta_{1,2,0} a + \delta_{2,1,0} \phi = 0\,\text{,}
\label{omegaprob}
\qqq
where the third is the commutativity of diagram \erf{17}, and the 2-form 
$\omega$ is regarded as an element in $A^2(\mathfrak{O}^2)$ corresponding 
to the trivial gerbe $\mathcal{I}_{\omega}$. Finally, the $Z^3$-equivariant 
2-isomorphism $\alpha$ has local data $d\in K^{3,0,0}$ such that
\qq
\Delta_{2,0,1} \beta = D_{3,0,0} d
\,\text{, }\quad
\Delta_{2,1,0} \phi + \delta_{3,0,0} d=0
\quad\text{ and }\quad
\Delta_{3,0,0}d=0\,\text{.}
\nonumber
\qqq
where the second is the commutativity of diagram \erf{35} and the third is  
the pentagon axiom for $\alpha$.

\smallskip

We also need to relate local data of two equivalent $Z$-multiplicative 
gerbes. Suppose the equivalence is expressed by a $Z$-equivariant isomorphism 
$(\mathcal{B},\kappa_z)$ and a $Z^2$-equivariant 2-isomorphism $\beta$ as 
discussed in Sec.\,\ref{sec:emgerbes}. Suppose further that we have chosen 
a $Z$-simplicial sequence of open covers that are fine enough to extract 
local data of all involved gerbes and isomorphisms. Then, the  
$Z$-equivariant isomorphism 
$(\mathcal{B},\kappa_z)$  has local data 
$r\in K^{1,0,1}$ and $s\in K^{1,1,0}$, and the  $Z^2$-equivariant 
2-isomorphism $\beta$  has local data 
$t\in K^{2,0,0}$. These relate local data
 $(c_1, b_1, a_1, \beta_1, \phi_1,d_1)$ and $(c_2, b_2, a_2, \beta_2, 
\phi_2,d_2)$ of the $Z$-multiplicative gerbes by
\qq
c_2 = c_1 + D_{1,0,1}r
\,\text{, }\quad
b_2=b_1 + \delta_{1,0,1}r + D_{1,1,0}s
\quad\text{ and }\quad
a_2=a_1 - \delta_{2,0,1}s\,\text{,}
\label{relA}
\qqq 
the last equation expressing the equivariance of $(\mathcal{B},\kappa_z)$, and 
\qq
\beta_2 = \beta_1 + \Delta_{1,0,1}r + D_{2,0,0}t
\,\text{, }\quad
\phi_2 = \phi_1  + \Delta_{1,1,0}s - \delta_{2,0,0}t\text{,}
\quad\text{ and }\quad
d_2=d_1 - \Delta_{2,0,0}t\,\text{.} 
\label{relB}
\qqq

We remark that the local data of a $Z$-multiplicative gerbe does not 
automatically define a cocycle in the total complex of the triple complex 
$K^{p,q,n}$, due to the appearance of the 2-form $\omega$ in \erf{omegaprob}. 
In \cite{WMult} the 2-form has been included into the complex, but here 
this will not be necessary.

\section{Obstructions against Equivariant Multiplicative Structures}
\label{sec:obs}

\no
Let $G$ be a 2-connected Lie group, and let $Z$ be a finite group acting 
smoothly and multiplicatively on $G$. Let $\mathcal{G}$ be a gerbe over $G$ 
of curvature $H$, and let $\omega\in\Lambda^2(G^2)$ 
be a multiplicative 2-form such that identity \erf{3Hnew} is satisfied. 
\smallskip

Due to the 2-connectedness of $G$, all the elements of a $Z$-multiplicative 
structure on $\mathcal{G}$ exist: the isomorphisms $\mathcal{A}_{z}$ of the 
equivariant structure on $\mathcal{G}$ and the isomorphism $\mathcal{M}$ over 
$G^2$ exist because they are isomorphisms between gerbes of equal curvature, 
and such gerbes are necessarily isomorphic. The 2-isomorphisms 
$\varphi_{z_1,z_2}$ of the equivariant structure on $\mathcal{G}$, the 
2-isomorphisms $\eta_{z_1,z_2}$ of the equivariant structure on $\mathcal{M}$, 
and the 2-isomorphism $\alpha$ of the multiplicative structure exist because 
every two isomorphisms between fixed gerbes are necessarily 2-isomorphic over 
simply connected spaces. \emph{Not} automatically guaranteed are the various 
conditions these 2-isomorphisms have to satisfy, namely the commutativity of 
the diagrams \erf{5}, \erf{17}, \erf{35} and the pentagon axiom for $\alpha$.

\smallskip

In order to handle all these conditions, let us extract local data for  the 
structure that we have chosen. We choose an appropriate $Z$-simplicial 
sequence $\left \lbrace \mathfrak{O}^p \right \rbrace$ as discussed 
in Sec.\,\ref{sec:local}.  According to the discussion there, we obtain 
local data $(c,b,a)$ for the $Z$-equivariant 
gerbe $\mathcal{G}$, local data $(\beta,\phi)$ for the $Z^2$-equivariant 
isomorphism $\mathcal{M}$ and local data $d$ for the 2-isomorphism $\alpha$.
All relations are satisfied except
\qq
\delta_{1,2,0} a=0
\,\text{, }\quad
\Delta_{1,2,0} a + \delta_{2,1,0} \phi = 0
\,\text{, }\quad
\Delta_{2,1,0} \phi + \delta_{3,0,0} d=0
\quad\text{ and }\quad
\Delta_{3,0,0}d=0\,\text{,}
\label{relations}
\qqq
corresponding to the above-mentioned diagrams and the pentagon axiom, 
respectively. These four equations are the cocycle condition for the 
3-cochain $(a,\phi,d)$ in the total complex of the double complex 
$K^{p,q,0}$ with differentials $\Delta_{p,q,0}$ and $\delta_{p,q,0}$. 
In the following we suppress the third index $0$.

\smallskip

In the generic case our chosen data does not satisfy the conditions 
\erf{relations}, and we define
\qq
u_0 := \delta_{1,2} a\in K^{1,3}
\text{, }\ \ 
u_1 := \Delta_{1,2} a + \delta_{1,2} \phi\in K^{2,2}
\text{, }\ \ 
u_2 := \Delta_{2,1} \phi - \delta_{3,0} d \in K^{3,1} 
\text{, }\ \ 
u_3 := \Delta_{3,0} d \in K^{4,0}\text{.}\ \quad
\nonumber
\qqq
Since this is the coboundary of $(a,\phi,d)$, the 4-cochain 
$u:=(u_0,u_1,u_2,u_3)$ is a 4-cocycle.
%, i.e.
%\qq
%\delta_{1,3}u_0=0
%\;\text{, }\quad
%\Delta_{1,3} u_0 - \delta_{2,2} u_1 =0
%\;\text{, }\quad
%\Delta_{2,2} u_1 - \delta_{3,1} u_2 =0
%\;\text{, }\quad
%\Delta_{3,1} u_2 + \delta_{4,0} u_3 =0
%\;\text{, }\quad
%\Delta_{4,0} u_3 =0\text{.}
%\qqq
By construction, $D_0u_i=0$ for all $i=0,...,3$. Now we recall the 
identification
\qq
\mathrm{ker}D_0|_{A^0(\mathfrak{O}^p)}  \cong U(1)
\label{identi}
\qqq
due to the fact that all manifolds $G^p$ are connected. 
We have thus induced identifications
\qq
\label{6}
K^{p,q,0} \supset\mathrm{ker}D_{p,q,0} \cong C^q(Z^p,U(1)) \text{,}
\qqq
where $C^q(Z^p,U(1))$ is the $q$-th cochain group of the group $Z^p$ with 
coefficients in $U(1)$, with $Z^p$ acting trivially on the coefficients. 
Unlike in Sec.\,\ref{sec:intro}-\ref{sec:cohFGK}, we shall use below the
additive notation rather than the multiplicative one for the $U(1)$-valued 
cochains. Under the identification 
\erf{6}, the remaining differentials $\delta_{p,q}$ and $\Delta_{p,q}$ are
\qq
\delta_{p,q}: C^q(Z^p,U(1)) \to C^{q+1}(Z^p,U(1))
\quad\text{ and }\quad
\Delta_{p,q}: C^q(Z^p,U(1)) \to C^q(Z^{p+1},U(1))\text{,}
\nonumber
\qqq
with $\delta_{p,q}$ the usual group cohomology differential for the group 
$Z^p$. The following identities are straightforward to check:
\begin{enumerate}
\item
$\delta_{p,0}=0$, since the action of $Z^p$ on the coefficients $U(1)$ 
is trivial,
\item 
$\Delta_{p,0}: U(1) \to U(1): z \mapsto \left\lbrace\begin{array}{ll}1 & 
\text{ for }p\text{ even,}  \\
z & \text{ for }p\text{ odd,} \\
\end{array}\right .$
\item
$\Delta_{p,1} = \delta_{1,p}$
under the  set-theoretic equality
$C^1(Z^p,U(1))=C^p(Z,U(1))$,

\item
$\Delta_{p,p}u=0$ if and only if $\delta_{p,p}u=0$.
\end{enumerate} 
The total cohomology of the double complex $C^q(Z^p,U(1))$ is denoted 
$\mathbb{H}_0^n(Z,U(1))$,
where the $0$ indicates that our double complex starts at $q = 0$ (but at 
$p=1$). 
The cocycle $u:=(u_0,...,u_3)$ we have obtained from the pair 
$(\mathcal{G},\omega)$ as described above is now a 4-cocycle and defines 
a class $[u]\in \mathbb{H}^4_0(Z,U(1))$. The cocycle conditions are \qq
\label{11}
\delta_{1,3}u_0=0
\quad\text{, }\quad
\Delta_{1,3} u_0 = \delta_{2,2} u_1 
\quad\text{, }\quad
\Delta_{2,2} u_1 = \delta_{3,1} u_2
\quad\text{ and }\quad
\Delta_{3,1} u_2  =0\text{.}
\qqq

Before we proceed investigating more closely the class 
$[u]\in\mathbb{H}^4_0(Z,U(1))$, let us make two general claims about it. 
The first is that $[u]$ is uniquely determined 
by the pair $(\mathcal{G},\omega)$ of the gerbe $\mathcal{G}$ and the 2-form 
$\omega$. The case that, for a fixed choice of 1-isomorphisms and 
2-isomorphisms, a different $Z$-simplicial open cover has been used can 
be reduced to the case that one cover refines the other. Then, local data 
for the finer cover can be chosen as the restriction of the local data 
for the coarser one. In this case, the identification \erf{identi} produces 
the same $U(1)$-numbers and hence the same cocycle $u$. Presume further 
that we either had chosen a different 
multiplicative or equivariant structure on $\mathcal{G}$, or chosen 
different local data. Due to the 2-connectedness of $G$
one can then find local data $(r,s,t)$ of an 
equivalence between $Z$-multiplicative gerbes, relating local data 
$(c,b,a,\beta,\phi,d)$ to other local data $(c',b',a',\beta',\phi',d')$ by
\qq
c_2 = c_1 + D_{1,0,1}r
\quad\text{, }\quad
b_2=b_1 + \delta_{1,0,1}r + D_{1,1,0}s
\quad\text{ and }\quad
\beta_2 = \beta_1 + \Delta_{1,0,1}r + D_{2,0,0}t\text{.}\quad
\nonumber
\qqq
These are Eqs.\,\erf{relA} and \erf{relB} minus the equations corresponding 
to the commutativity of diagrams of 2-isomorphisms, which are not 
automatically guaranteed. Anyway, it is now easy to see that
\qq
x := a'-a + \delta_{1,1}s
\quad\text{, }\quad
y:= \phi' - \phi - \Delta_{1,1}s + \delta_{2,0}t
\quad\text{ and }\quad
z:= d'-d + \Delta_{2,0}t
\nonumber
\qqq
defines a 3-cochain $(x,y,z)$ in the total complex of $C^q(Z^p,U(1))$, whose 
coboundary is $u'-u$. Thus, the class $[u]$ is well-defined. 

\smallskip

The second claim about the class $[u]$ is that it is the  obstruction against 
the existence of a $Z$-multiplicative structure for the pair 
$(\mathcal{G},\omega)$, i.e. there exists a $Z$-multiplicative structure 
if and only if $[u]=0$. The ``only if'' part is trivial: if there is a 
$Z$-multiplicative structure on $(\mathcal{G},\omega)$, the corresponding 
cocycle $u$ is identically zero, since all the relations \erf{relations} 
are satisfied. Conversely, suppose  $(x,y,z)$ is a 3-cochain whose coboundary 
is $(u_0,...,u_4)$. Then, the new local data $(c,b,a-x,\rho,\beta,\phi-y,d-z)$ 
for a $Z$-multiplicative gerbe satisfies all required conditions. 
Reconstructing the gerbe, isomorphisms and 2-isomorphisms from this local  
data yields a $Z$-multiplicative gerbe with the underlying gerbe isomorphic to 
$\mathcal{G}$ and with the 2-form $\omega$. 
%KG: I added the following phrase:
The latter isomorphism allows to carry the $Z$-multiplicative structure
to $\mathcal{G}$.

\smallskip

Now we analyze the obstruction class $[u]\in \mathbb{H}^4_0(Z,U(1))$ in 
detail. As the cocycle conditions
\erf{11} show, the obstructions split into one obstruction $u_3\in U(1)$ 
and a class $[(u_0,u_1,u_2)]$ in $\mathbb{H}^4(Z,U(1))$, the total cohomology 
of the double complex $C^q(Z^p,U(1))$ with the $(q=0)$-row omitted. Since the 
differential
$\Delta_{3,0}$
is the identity, we can always find a cohomologous 4-cocycle with $u_3=0$. 
And because the differential
$\delta_{3,0}$
is the zero map, the class $[(u_0,u_1,u_2)] \in \mathbb{H}^4(Z,U(1))$ is 
trivial if and only if the class $[u]\in \mathbb{H}^4_0(Z,U(1))$ is trivial. 
Summarizing, there is one well-defined obstruction $[(u_0,u_1,u_2)] \in 
\mathbb{H}^4(Z,U(1))$ against a $Z$-multiplicative structure on 
$(\mathcal{G},\omega)$.

\smallskip

The obstruction class $[u]$  can further be  treated as follows. Since 
$\delta_{1,3}u_0=0$, we have a class 
$[u_0]\in H^3(Z,U(1))$. Suppose this class is trivial and that 
$x\in C^2(Z,U(1))$ is such that $\delta_{1,2}x=u_0$.  If we now change 
$u$ by the coboundary of $(x,0)$, we obtain a new representative 
$u^x:=(0,u_1^x,u_2)$ with $u_1^x:=u_1 - \Delta_{1,2} x$, in which 
$\delta_{2,2}u^x_1=0$.  
It defines a class $[u_1^x] \in H^2(Z^2,U(1))$.
Note that this class depends on the choice of $x$. Now suppose that the 
class $[u_1^x]$ is trivial, i.e. there exists $y\in C^1(Z^2,U(1))$ such 
that $\delta_{2,1}y = u_1^x$. We change $u^x$ be the coboundary of $(0,y)$ 
and obtain a new representative $u^{x,y}=(0,0,u^{y}_2)$ with $u^{y}_2 
:= u_2 - \Delta_{2,1}y$. The cocycle conditions are now $\delta_{3,1}u_2^y=0$
and $\Delta_{3,1}u_2^y=0$. 
%KG: I expanded what follows 
This relations means that $u^y_2$ is a character
on $Z^3$ that is also a 3-cocycle if viewed as a 3-chain on $Z$.
A short computation shows that $\,(u^y_2)_{(z,z',z'')}=\chi(z)+\chi'(z'')$,
\,where $\,\chi$ and $\chi'$ are characters on $Z$. Upon setting
$v_(z,z')=-\chi(z)+\chi'(z')$, one checks that   
$v\in C^1(Z^2,U(1))$ satisfies $\Delta_{2,1}v=u^y_2$ and $\delta_{2,1}v=0$.
Altogether, this implies that the cocycle $u=(u_0,u_1,u_2)$ is a coboundary 
of $(x,y+v)$. Thus, the obstruction $[u]\in\mathbb{H}^4(Z,U(1))$ is trivial 
if, and only if, successively, the class of $u_0$ in $H^3(Z,U(1))$ and the 
class of $u_1^x$ in $H^2(Z^2,U(1))$ vanish. 
\smallskip

Tracing back through the extraction of local data, we see that $u_0$ is 
the error in the commutativity of the diagram \erf{5} for the 2-isomorphism 
$\varphi_{z_{1},z_2}$ of equivariant structures. Thus, its class 
$[u_0] \in H^3(Z,U(1))$ is the well-known obstruction from \cite{GR04,GSW}
to the existence of the descent gerbe $\mathcal{G}'$ on $G'=G/Z$. 
Further, once an equivariant structure (the local datum $x$) is chosen, 
we see that $u_1^x$ is the error in the commutativity of the diagram which 
is needed to make $(\mathcal{M},\eta_{z_1,z_2})$ a $Z^2$-equivariant 
isomorphism. Thus, the class $[u_1^x]\in H^2(Z^2,U(1))$ obstructs the 
existence of the descent isomorphism $\mathcal{M}'$ in the multiplicative 
structure on the gerbe $\mathcal{G}'$.

\smallskip

For the particular case of $\mathcal{G}_k$ the basic gerbe over a compact, 
simple and simply-connected Lie group $G$, and $\omega_k$ the 2-form 
\erf{rho}, we have shown in Sec.\,\ref{sec:cohFGK}, that the class 
$\kappa = [u_1^x]$ 
is trivial if and only if the FGK cocycle $c$ associated to $\mathcal{G}_k'$ is
trivial (note that the groups $G$ and $G'=G/Z$ play here the role of  
$\tilde G$ and $G=\tilde G/Z$ from the first sections of the paper
where the discussion was centered on the gerbes over non-simply connected 
groups). Thus, the calculation of the FGK cocycle carried out 
in Sec.\,\ref{sec:calFGK} identifies precisely the 
the situations for which $\mathcal{G}_k'$ is a multiplicative gerbe.

\section{Uniqueness of Multiplicative Structures}
\label{sec:unique}

\noindent In this section we address the  question 
if there are inequivalent choices 
of $Z$-multiplicative structures on a pair $(\mathcal{G},\omega)$ of a gerbe 
$\mathcal{G}$ over a 2-connected Lie group $G$ and a compatible 2-form 
$\omega$. First we claim that equivalence classes of $Z$-multiplicative 
structures on $(\mathcal{G},\omega)$, if they exist, are parameterized by 
$\mathbb{H}^3(Z,U(1))$. 
\smallskip

Let us first see how $\mathbb{H}^3(Z,U(1))$ acts on equivalence classes of 
$Z$-multiplicative gerbes. A 3-cocycle consists of cochains 
$x\in C^2(Z,U(1))$ and  $y\in C^1(Z^2,U(1))$ such that 
\qq
\delta_{1,2}x=0
\quad\text{, }\quad
\Delta_{1,2}x + \delta_{2,1}y =0
\quad\text{ and }\quad
\Delta_{2,1}y=0\text{.}
\nonumber
\qqq
If $(c,b,a,\beta,\phi,d)$ is local data for a $Z$-multiplicative gerbe then 
$(c,b,a+x,\beta,\phi+y,d)$ is local data for another one. Suppose that
$(x',y')$ is a cohomologous cocycle, i.e. there exists $s \in C^1(Z,U(1))$ 
such that $x'=x+\delta_{1,1}s$ and $y'=y+\Delta_{1,1}s$. Then, 
$(0,s,0)$ is local 
data for an equivalence between $(c,b,a+x,\beta,\phi+y,d)$ and 
$(c,b,a+x',\beta,\phi+y',d)$. Thus, the action of $\mathbb{H}^3(Z,U(1))$ 
on equivalence classes of $Z$-multiplicative gerbes is well-defined. It is 
also free: whenever $(c,b,a,\beta,\phi,d)$ and $(c,b,a+x,\beta,\phi+y,d)$ 
are equivalent local data for an equivalence $(r,s,t)$, one can choose 
$r=t=0$ and verify that the coboundary of $s$ is $(x,y)$. It remains to show 
that the action is transitive.

If $(c,b,a,\beta,\phi,d)$ and $(c',b',a',\beta',\phi',d')$ are local data 
for $Z$-multiplicative gerbes over $G$, with $c$ and $c'$ local data for 
the fixed gerbe $\mathcal{G}$, one can always find local data $(r,s,t)$ of 
an equivalence that change the second to $(c,b,a'',\beta,\phi'',d'')$, as 
mentioned in Sec.\,\ref{sec:obs}. One can also arrange $t$ such that $d''=d$, 
due to the fact that multiplicative structures on $\mathcal{G}$ are unique 
\cite{WMult}. Defining $x := a''-a$ and $y:= \phi''-\phi$, we obtain a class 
$[(x,y)] \in \mathbb{H}^3(Z,U(1))$  whose action connects the two sets of 
local data. Thus, the action of $\mathbb{H}^3(Z,U(1))$ is transitive. 

\smallskip

We have shown so far that $\mathbb{H}^3(Z,U(1))$ parameterizes inequivalent 
$Z$-multiplicative structures on a gerbe $\mathcal{G}$ over a 2-connected 
Lie group $G$. Now we calculate $\mathbb{H}^3(Z,U(1))$ for the groups 
$Z=\NZ_N$ and $Z=\NZ_2^2$ that appear as subgroups of centers 
of compact, simple and simply-connected Lie groups. We first discuss the 
case of $Z = \NZ_N$, for which we consider a 3-cocycle $(x,y)$. Since 
$H^2(\NZ_N,U(1))=0$ according to \erf{ucth0}, we can go to a representative 
with $x=0$. Then, 
$\delta_{2,1}y=0$ and $\Delta_{2,1}y=0$, 
%KG: I expanded a little:
i.e.\,$y\in C^1(Z^2,U(1))$ is a character on $Z^2$ that is also a 
2-cocycle when viewed as a 2-chain on $Z$. A simple calculation 
shows that $y=0$. 
Thus, $\mathbb{H}^3(\NZ_N,U(1))=0$. 
We continue with $Z = \NZ_2^2$. Here, $H^2(Z,U(1))=\NZ_2$, see \erf{ucth0}. 
One can represent 
the non-trivial class explicitly.  Under the embedding \erf{emb},
\qq
\mathrm{Hom}(\NZ_2,U(1)) \cong \mathrm{Hom}(\NZ_2 \otimes \NZ_2,U(1)) 
\hookrightarrow H^2(\NZ_2^2,U(1))\text{,}  
\nonumber
\qqq
the non-trivial group homomorphism $\zeta:\NZ_2 \to U(1)$ maps to the 
non-trivial class represented by $\chi_{(z_1,z_2),(z_1',z_2')} := 
\zeta(z_1z_2')$. We can thus  check explicitly that
\qq
(\Delta_{1,2}\chi)_{((z_1,z_2),(\tilde z_1,\tilde z_2)),
((z_1',z_2'),(\tilde z_1',
\tilde z_2'))} = 0\text{.}
\nonumber
\qqq
It follows as before that $y=0$. 
Consequently,
$\mathbb{H}^3(\NZ_2^2,U(1))=\NZ_2$. This is just the well-known 
choice of the $\NZ_2^2$-equivariant structure on the gerbe $\mathcal{G}_k$ 
over $Spin(4r)$ \cite{GR04,GSW}. 

\smallskip

Summarizing, once an equivariant structure on the gerbe $\mathcal{G}_k$ over 
$G$ is fixed, the multiplicative structure on the descent gerbe 
$\mathcal{G}_k'$ is, if it exists, unique
up to isomorphism.

\section{Conclusions}

\label{sec:concl}

\no We have studied obstructions to the existence of multiplicative structures
on (bundle) gerbes $\,\CG_k\,$ over simple compact groups $\,G$, 
\,i.e.\,\,on gerbes with connection of curvature $\,H_k\,$ given by 
Eq.\,(\ref{meg}). This was done by analyzing the multiplicative gerbes over 
the universal covering groups $\,\tilde G\,$ equivariant w.r.t. the deck 
action of the fundamental group $\,\pi_1(G)=Z$. \,We have shown that there 
are two obstructions to the existence of such equivariant multiplicative
gerbes over $\,\tilde G$. \,The first one lies in the cohomology groups $\,H^3(Z,U(1))$. 
\,Its triviality assures the existence of the gerbe $\,\CG_k\,$ over the 
quotient group $\,G=\tilde G/Z$. \,Such a gerbe determines unambiguously 
the Feynman amplitudes in the group $\,G\,\,$ WZW theory over closed 
oriented surfaces. Given the gerbe $\,\CG_k\,$ over $\,G$, \,the
second obstruction lies in the cohomology group $\,H^2(Z^2,U(1))$.
\,Its triviality, equivalent to the triviality of the FGK 2-cocycle 
on the group $\,Z^2$,  guarantees the existence of the multiplicative 
structure on the gerbe $\,\CG_k$. \,The gerbe $\,\CG_k$ with such a
structure determines unambiguously the Feynman amplitudes in the group 
$\,G\,\,$ CS theory over closed oriented 3-dimensional manifolds.
We made explicit the relation between the obstruction cohomology class in     
$\,H^2(Z^2,U(1))\,$ and the FGK cocycle by relating both to Kreuzer-Schellekens
bihomomorphisms. 
\,The constraints on the integer levels $\,k\,$ imposed 
by the triviality of, respectively, the first obstruction or the first 
and the second one are collected in the table in Fig.\,2. 
\begin{figure}[ht]
\begin{center}
\begin{tabular}{|c|c|c|c|c|}\hline
$\tilde G$&Center&$Z$&WZW constraints on $k$&CS constraints on $k$\\\hline
$SU(r)$ & $\NZ_{r}$ & $Z = \NZ_N$ with $N \;|\; n$ & $2N \;|\; 
kr(r-1)$  &$2N^2 \;|\; 
kr(r-1)$ \\\hline
$Spin(2r+1)$ & $\NZ_2$ & $Z=\NZ_2$ &$-$  &$2 \;|\; k$ \\\hline
$Spin(4r+2)$ & $\NZ_4$ & \begin{tabular}{c}$Z=\NZ_2$\\$Z=\NZ_4$\end{tabular}
& \begin{tabular}{c}$-$\\$2 \;|\; k$\end{tabular} 
& \begin{tabular}{c}$2 \;|\; k$\\$8 \;|\; k$\end{tabular} \\\hline
$Spin(4r)$ & $\NZ_2\times\NZ_2$ & \begin{tabular}{c} $Z = \NZ_2\times\{0\}$  
\\ $Z = \{0\}\times\NZ_2$\\ $Z = \{(0,0),(1,1)\}$ \\ $Z= \NZ_2\times \NZ_2$
\end{tabular}& 
\begin{tabular}{c} $2 \;|\; kr$\\$-$\\$2 \;|\; kr$\\$2 \;|\; kr$
\end{tabular} & 
\begin{tabular}{c} $4 \;|\; kr$\\$2 \;|\; k$\\$4 \;|\; kr$\\$2\;|\;k\ \text{and}
\ 4 \;|\; kr$
\end{tabular} \\\hline
$Sp(2r)$ & $\NZ_2$ & $Z=\NZ_2$ & $2 \;|\; kr$ & $4 \;|\; kr$ \\\hline
$E_6$ & $\NZ_3$ & $Z=\NZ_3$ & $-$ & $3 \;|\; k$ \\\hline
$E_7$ & $\NZ_2$ & $Z=\NZ_2$ & $2 \;|\; k$ & $4 \;|\; k$ \\\hline
\end{tabular}
\end{center}
\caption{\ Constraints for integer values of $k$ imposed by the consistency
of, \,respectively, \,the WZW and the CS theory\\ \hspace*{1.25cm}with group 
$\,G=\tilde G/Z$}
\label{tab1}
\end{figure}
Over 
the groups $\,G=Spin(4r)/\NZ_2^2\,$ where there are two inequivalent gerbes $\,\CG_k\,$ when $\,kr\,$ is even, only one of them admits multiplicative
structure when $\,kr\,$ is divisible by $\,4$. 
For simplicity, we have limited our geometric considerations to the case of simple compact groups. The extension of our analysis to the case of 
non-simple compact groups, treated within the simple-current orbifold 
algebraic approach in \cite{SY,SY1,GRS,KS,FRS} does not seem to present 
difficulties. 
\vskip 0.1cm
  
Multiplicative structures on gerbes over groups may be viewed as conditions 
of compatible equivariance of gerbes under the group actions on itself by 
the left and the right multiplications. We shall develop
elsewhere a theory of gerbes equivariant under actions of continuous
groups. Such gerbes permit the treatment of sigma models with 
the gauged Wess-Zumino actions, e.g.\,\,the coset models of conformal
field theory \cite{GK}. One of the useful applications of the
multiplicative structures on the gerbe $\,\CG_k\,$ over group $\,G\,$      
is that such a structure induces an equivariant structure w.r.t. 
the adjoint action of $\,G\,$ on itself (although the latter exists also 
when there is no multiplicative structure on $\,\CG_k$). \,A multiplicative 
structure on gerbes $\,\CG_k\,$ induces also two Jandl structures on 
$\,\CG_k\,$ (by the pullback of 1-isomorphisms $\,\CM\,$ of (\ref{multmorph}) 
from $\,G^2\,$ to $\,G\,$ via the maps $\,g\mapsto(g,g^{-1})\,$ or 
$\,g\mapsto(g^{-1},g)$. \,Such structures are used to define the amplitudes 
of the WZW theory over unoriented surfaces \cite{SSW,GSW}. Finally, 
multiplicative structure on gerbes $\,\mathcal{G}_k\,$ play an important 
role in WZW theory with defects permitting to define symmetric 
bi-branes \cite{FSW,WMult}. A detailed study of such defects for non-simply 
connected groups is another problem left for future research. 
\vskip 0.1cm

\section*{Appendix: Simplicial and Equivariant Refinements of Sequences of 
Open Covers}

Let $G$ be a Lie group, and let $Z$ be a finite Abelian group with a 
smooth, free, proper and multiplicative action on $G$. Suppose that 
$\left \lbrace \mathfrak{V}^p \right \rbrace$ is a sequence consisting 
of open covers $\mathfrak{V^p}$ of $G^p$ for $p>0$. In this appendix we 
show that there exists a $Z$-simplicial sequence $\left \lbrace 
\mathfrak{O}^p \right \rbrace$ open covers in the sense of 
Sec.\,\ref{sec:local}, such that each $\mathfrak{O}^p$ is a refinement 
of $\mathfrak{V}^p$.  

\smallskip

\def\set{\mathcal{S}\!et}
\def\man{\mathcal{M}\!an}

We have to introduce some simplicial techniques. Let $\mathcal{P}$ be the 
category whose objects are non-negative integer numbers $0,1,2,...$ 
and whose set $\mathcal{P}(n,m)$ of morphisms from $n$ to $m$ consists 
of maps $\theta: \left \lbrace 0,...,n \right \rbrace \to \left \lbrace 
0,...,m \right \rbrace$ such that $i<j$ implies $\theta(i)<\theta(j)$. 
If $n>m$, the set $\mathcal{P}(n,m)$ is empty. If $n=m$, it contains only 
the identity. Furthermore, for any $n>0$ the set $\mathcal{P}(n-1,n)$ consists 
of $(n+1)$ elements, the ``universal face maps'' $\theta_k$, where 
$0\leq k \leq n$, defined by $\theta_k(i) := i$ for $i<k$ and $\theta_k(i) 
:=i+1$ for $i\geq k$. The category $\mathcal{P}$ has the following well-known 
purpose. There is a bijection between incomplete simplicial sets and 
contravariant functors $X: \mathcal{P} \to \set$, where $\set$ denotes the  
category of sets. Given such a functor, one obtains an incomplete simplicial 
set $\left \lbrace X^p \right \rbrace$ by setting $X^p := X(p)$ and 
$\Delta^p_k := X_{\theta_k} := X(\theta_k)$ for $\theta_k\in \mathcal{P}
(p-1,p)$ one of the universal face maps. Conversely, one can write any 
$\theta \in \mathcal{P}(n,m)$ as a composition of universal face maps and 
then invert this construction. In particular, every group $G$ defines a 
contravariant functor $G: \mathcal{P} \to \set$, corresponding to the 
incomplete simplicial set $\left \lbrace G^p \right \rbrace$ that we 
considered at the beginning of Sec.\,\ref{sec:emgerbes}. An analogous 
statement is true in the category $\man$ of smooth manifolds: 
the contravariant functors $M: \mathcal{P} \to \man$ are in bijection 
to incomplete simplicial manifolds. 

\smallskip

First we recall a general construction of \cite{Tu}.
Let $M: \mathcal{P} \to \mathcal{M}\!an$ be a contravariant functor and 
let $\left \lbrace \mathfrak{V}^p \right \rbrace$ be a sequence  of open 
covers, with $\mathfrak{V}^p =  \lbrace V^p_{j}  \rbrace_{j \in J^p}$  an 
open cover of the manifold $M^p$. In the following we use the notation 
\qq
\mathcal{P}^p := \bigcup_{k=0}^{p} \mathcal{P}(k,p)
\quad\text{ and }\quad
\mathcal{J}^p := \bigcup_{k=0}^p J^k\text{.}
\nonumber
\qqq
A new open cover $\mathfrak{O}^p$ of $M^p$ is defined as follows. Its index 
set is
\qq
I^p := \left \lbrace i: \mathcal{P}^p \to \mathcal{J}^p \;\;|\;\;  
i(\theta) \in J^k \text{ for  }\theta \in \mathcal{P}(k,p)  \right 
\rbrace\text{.}
\nonumber
\qqq
Its open sets are
\qq
O^p_i := \bigcap_{k=0}^p \;\; \bigcap_{\theta \in \mathcal{P}(k,p)}\;\; 
M_\theta^{-1}(V^k_{i(\theta)})\text{,}
\label{defopenset}
\qqq
where $M_{\theta}: M^p \to M^k$ is the smooth map assigned to $\theta$.
These open sets cover $M^p$: for $x\in M^p$ choose an index 
$j_{\theta}\in J^k$ for each $\theta\in \mathcal{P}(k,p)$ such that 
$M_\theta(x) \in V^k_{j_{\theta}}$. The assignment $\theta \mapsto j_{\theta}$ 
defines an index $i \in I^p$, and it is clear that $x \in O^p_i$. Furthermore, 
$\mathfrak{O}^p$ is a refinement of $\mathfrak{V}^p$: with $r: I^p \to J^p$ 
defined by $r(i) := i(\id_p)$ we have immediately $O_i^p \subset V^p_{r(i)}$. 

Next we define a contravariant functor $I: \mathcal{P} \to \mathcal{S}\!et$ 
with $I(p) := I^p$, turning the sequence $\left \lbrace I^p \right \rbrace$ 
of index sets into an incomplete simplicial set. For $\phi \in \mathcal{P}
(n,m)$, we let $I_{\phi} : I^m \to I^n$ be defined by 
\qq
I_\phi(i)(\theta) := i(\phi \circ \theta) \in J^k
\nonumber
\qqq
for $i\in I^m$ and $\theta \in \mathcal{P}(k,n)$. It is clear that this 
definition yields a functor. Now we have all the structure of a simplicial 
sequence of open covers, and it remains to check condition \erf{simpcond}. 
In fact the more general relation
\qq
M_\theta(O_i^p)  \subset O^{\ell}_{I_\theta(i)}
\label{gensimpcond}
\qqq
is true for all $i\in I^p$ and $\theta \in \mathcal{P}(\ell,p)$, from 
which \erf{simpcond} follows by restricting to the universal face maps 
$\theta_k \in \mathcal{P}(p-1,p)$. To show \erf{gensimpcond} we have to 
prove that $M_\theta(O_i^p)$ is contained in all the open sets 
$M_{\phi}^{-1}(V^{k}_{I_\theta(i)(\phi)})$ that form the intersection 
\erf{defopenset}. Indeed,
\qq
M_\phi(M_\theta(O_i^p)) = M_{\theta \circ \phi}(O_i^p)\ \subset\ 
\bigcap_{k=0}^p \;\; \bigcap_{\psi \in \mathcal{P}(k,p)}\;\; M_{\theta 
\circ \phi}(M_\psi^{-1}(V^k_{i(\psi)}))\ \subset\ V^k_{i(\theta \circ \phi)} 
=V^{k}_{I_\theta(i)(\phi)} \text{.} 
\nonumber
\qqq
Here, the last inclusion follows by restricting to $\psi = \theta \circ 
\phi$. Summarizing the construction we took from \cite{Tu}, the sequence 
$\left \lbrace \mathfrak{O}^p \right \rbrace$ of open covers is simplicial, 
and each cover $\mathfrak{O}^p$ is a refinement of the original open cover 
$\mathfrak{V}^p$.

\smallskip 

Now we enhance the construction above by additional $Z$-equivariance. 
Since we have a free and proper group action of a finite group it is 
clear that each open cover $\mathfrak{V}^p$ has a $Z^p$-equivariant 
refinement. We can thus assume that such refinements are already chosen, 
and that the sequence $\left \lbrace \mathfrak{V}^p \right \rbrace$ is 
$Z$-equivariant. It remains to prove that applying the above construction 
to a $Z$-equivariant sequence $\left \lbrace \mathfrak{V}^p \right \rbrace$ 
yields a $Z$-simplicial sequence. To start with, the action of $Z^p$ on 
the index set $I^p$ is defined by 
\qq
(z.i)(\theta) := Z_\theta(z).i(\theta)\text{,}
\label{newaction}
\qqq
for $\theta \in  \mathcal{P}(k,p)$ and $z\in Z^p$. Here $Z_\theta: Z^p \to 
Z^k$ is the map the functor $Z: \mathcal{P} \to \set$ associated to the 
group $Z$ assigns to $\theta$, and on the right hand side we have used 
the action of $Z_\theta(z)\in Z^k$ on the index $i(\theta) \in J^k$ of 
the $Z^k$-equivariant open cover $\mathfrak{V}^p$. Definition \erf{newaction} 
yields an action because $Z_{\theta}$ is a group homomorphism; here 
we use that $Z$ is Abelian.
Notice that the refinement maps $r:I^p \to J^p$ defined above are 
$Z^p$-equivariant. Next we prove the relation \erf{equivcond} for 
$Z$-equivariant covers: first we have
\qq
z(O_i^p)\ \subset\ \bigcap_{k=0}^p \;\; \bigcap_{\theta \in \mathcal{P}(k,p)}
\;\; z \left ( M_\theta^{-1}(V^k_{i(\theta)})\right )\ \subset\ 
\bigcap_{k=0}^p \;\; \bigcap_{\theta \in \mathcal{P}(k,p)}\;\;
M_\theta^{-1}(Z_\theta(z)(V^k_{i(\theta)}))\text{,}
\nonumber
\qqq
for $z \in  Z^k$. Here we have used that the action is multiplicative; 
more specifically that diagram \erf{multact} is commutative. Then it 
remains to check that
\qq
 M_\theta^{-1}(Z_\theta(z)(V^k_{i(\theta)}))\subset M_\theta^{-1}
(V^{k}_{Z_\theta(z).i(\theta)}) = M_\theta^{-1}(V^{k}_{(z.i)(\theta)})\text{.}
\nonumber
\qqq
This shows that the sequence $\left \lbrace \mathfrak{O}^p \right 
\rbrace$ of open covers is $Z$-equivariant. It remains to check the 
compatibility condition \erf{multact1} between the face maps of 
$\left \lbrace I^p \right  \rbrace$ and the actions of $Z^p$ on $I^p$. 
Indeed, for $\theta \in \mathcal{P}(p-1,p)$,  $\phi \in \mathcal{P}(k,p-1)$, 
$z\in Z^p$ and $i\in I^p$ we find
\qq
(I_\theta(z.i))(\phi) = (z.i)(\theta \circ \phi) = Z_{\theta \circ 
\phi}(z).i(\theta \circ \phi) = Z_\phi(Z_\theta(z)).(I_\theta(i)(\phi))  
= (Z_\theta(z).I_\theta(i))(\phi)\text{,}
\nonumber
\qqq
showing the commutativity of \erf{multact1}. Summarizing, $\left \lbrace 
\mathfrak{O}^p \right \rbrace$ is a $Z$-simplicial sequence of open covers.


\begin{thebibliography}{bib}


\bibitem{AbGep}
{\sc A.~Abouelsaood and D.~Gepner}:
``Soliton strings, the WZW model and modular invariance'',
Phys. Lett. {\bf B 176} (1986), 380-386  

\bibitem{Brown}
{\sc K.~S.~Brown}: Cohomology of Groups, Springer, Berlin-Heidelberg-New York
1982

\bibitem{CMM}
{\sc A.~L.~Carey, J.~Mickelsson and M.~K.~Murray}: ``Bundle gerbes applied
to Quantum Field Theory'', Rev. Math. Phys. {\bf 12} (2000) 65-90

\bibitem{CJMSW}
{\sc A.~L.~Carey, S. Johnson, M.~K.~Murray, D. Stevenson and
B.~L.~Wang}: ``Bundle gerbes for Chern-Simons and Wess-Zumino-Witten
theories'', Commun. Math. Phys. {\bf 259} (2005), 577-613  

\bibitem{DijkWitt}
{\sc R.~Dijkgraaf and E.~Witten}: ``Topological gauge theories and 
group cohomology'', Comm. Math. Phys. {\bf 129} (1990), 393-429 

\bibitem{FSh}
{\sc L.~D.~Faddeev and S.~L.~Shatashvili}: ``Algebraic and Hamiltonian 
methods in the theory of non-Abelian anomalies'', Teor. Mat. Fiz.
{\bf 60} (1984), 206-217

\bibitem{FGK}
{\sc G.~Felder, K.~Gaw\c{e}dzki and A.~Kupiainen}: ``Spectra 
of Wess-Zumino-Witten models with arbitrary simple groups'', 
Commun. Math. Phys. {\bf 117} (1988), 127-158 

\bibitem{FRS}
{\sc J.~Fuchs, I.~Runkel and Ch.~Schweigert}:
``TFT construction of RCFT correlators III: Simple currents'',
Nucl. Phys. {\bf B 694} (2004), 277-353

%KG: I added the reference below
\bibitem{FSW}
{\sc J.~Fuchs, Ch.~Schweigert and K. Waldorf}:
``Bi-branes: target space geometry for world sheet topological defects'',
J. Geom. Phys. {\bf 58} (2008), 576-598 

\bibitem{GRS}
{\sc B.~Gato-Rivera and A.~N.~Schellekens}: ``Complete classification of
simple current automorphisms'', Nucl. Phys. {\bf B 353} (1991), 519-537

\bibitem{Gaj}
{\sc P.~Gajer}: ``Geometry of Deligne cohomology'', Invent. Math. {\bf 127}
(1997), 155-207

\bibitem{top}
{\sc K.~Gaw\c{e}dzki},
``Topological actions in two-dimensional quantum field theory'',
in: Non-perturbative Quantum Field Theory, G.~'t Hooft, A.~Jaffe,
G.~Mack, P.~Mitter and R.~Stora (eds.), Plenum Press, 1988, pp. 101-142

\bibitem{G90}
{\sc K.~Gaw\c{e}dzki}, ``Geometry of Wess-Zumino-Witten models 
of conformal field theory'', in: Recent Advances in Field Theory, eds. 
P.~Bin\'{e}truy, G.~Girardi, P.~Sorba, Nucl. Phys. (Proc. Suppl.) 
{\bf B18} (1990), 78-91

\bibitem{G05}
{\sc K.~Gaw\c{e}dzki}:
``Abelian and non-Abelian branes in WZW models and gerbes'',
Commun. Math. Phys. {\bf  258} (2005), 23-73

\bibitem{GK}
{\sc K.~Gaw\c{e}dzki and A. Kupiainen}: ``Coset construction from functional 
integrals, Nucl. Phys. {\bf B 320} (1989), 625-668

\bibitem{GR02}
{\sc K.~Gaw\c{e}dzki and N.~Reis}:
``WZW branes and gerbes'',
Rev.Math.Phys. {\bf 14} (2002), 1281-1334  

\bibitem{GR04} 
{\sc K.~Gaw\c{e}dzki and N.~Reis}, 
``Basic gerbe over non simply connected compact groups'', 
J. Geom. Phys. {\bf 50} (2004), 28-55 

\bibitem{GSW}
{\sc K.~Gaw\c{e}dzki, R.~R.~Suszek and K.~Waldorf}: ``Bundle gerbes for 
orientifold sigma models'', arXiv:0809.5125

\bibitem{GepWitt}
{\sc D.~Gepner and E.~Witten}: ``String theory on group manifolds'', 
Nucl. Phys. {\bf B 278} (1986), 493-549 

\bibitem{KS}
{\sc M.~Kreuzer and A.~N.~Schellekens}:
``Simple currents versus orbifolds with discrete torsion - a complete 
classification'', Nucl.Phys. {\bf B 411} (1994), 97-121 

\bibitem{ZOO}
{\sc G.~Moore and N.~Seiberg}: ``Taming the conformal zoo'',
Phys. Lett. {\bf B 220} (1989), 422-430

\bibitem{Murr} 
{\sc M.K.~Murray}: 
``Bundle gerbes'', 
J. London Math. Soc.\,(2) {\bf 54} (1996), 403-416 

\bibitem{MurrS} 
{\sc M.K.~Murray and D.~Stevenson}: 
``Bundle gerbes: stable isomorphisms and local theory'', 
J. London Math. Soc.\,(2) {\bf 62} (2000), 925-937 

\bibitem{McL}
{\sc S.~Mac Lane}: Homology, Springer, Berlin-Heidelberg-New York
1975

\bibitem{PW}
{\sc A.~M.~Polyakov and P.~B.~Wiegmann}: ``Goldstone fields in two dimensions 
with multivalued actions '', Phys. Lett. {\bf B 141} (1984), 223-228 

\bibitem{SY}
{\sc A.~N.~Schellekens and S.~Yankielowicz}:
``Extended chiral algebras and modular invariant partition functions'',
Nucl. Phys. {\bf B 327} (1989), 673-703

\bibitem{SY1}
{\sc A.~N.~Schellekens and S.~Yankielowicz}:
``Modular invariants from simple currents: an explicit proof'',
Phys. Lett. {\bf B 227} (1989), 387-391

\bibitem{SSW}
{\sc U.~Schreiber, Ch.~Schweigert and K.~Waldorf}: ``Unoriented WZW models 
and holonomy of bundle gerbes'', Commun. Math. Phys. {\bf 274} (2007), 31-64

\bibitem{Ste00}
{\sc D.~Stevenson}: ``The geometry of bundle gerbes'', PhD thesis,
University of Adelaide, 2000, arXiv:math.DG/0004117 

\bibitem{Tu}
{\sc J.-L.~Tu}: ``Groupoid cohomology and extensions'', Trans. Amer. 
Math. Soc. {\bf 358} (2006), 4721-4747

\bibitem{WMult}
{\sc K.~Waldorf}: ``Multiplicative bundle gerbes with connection'', Differential Geom. Appl., to appear,
arXiv:0804.4835 

\bibitem{Witt}
{\sc E.~Witten}: ``Non-abelian bosonization in two dimensions'',
Commun. Math. Phys. {\bf 92} (1984), 455-472

\bibitem{WittCS}
{\sc E.~Witten}: ``Quantum field theory and Jones polynomial'',
Commun. Math. Phys. {\bf 121} (1989), 351-399

\end{thebibliography}
\end{document}